\documentclass[pra,aps,floatfix,10pt,twocolumn,superscriptaddress,nofootinbib]{revtex4-2}

\usepackage[T1]{fontenc}
\usepackage[utf8]{inputenc}
\usepackage{microtype}

\usepackage{graphicx}
\usepackage{dcolumn}
\usepackage{bm}
\usepackage{amsmath}
\usepackage{amsfonts}
\usepackage{amssymb}
\usepackage{amstext}   
\usepackage{amsthm}
\usepackage[english]{babel}
\usepackage{makecell}
\usepackage{dsfont}
\usepackage{nameref}
\usepackage{color}
\usepackage{float}
\usepackage[outdir=./]{epstopdf}
\usepackage{nameref}
\usepackage{subfloat}
\usepackage[colorlinks = true, citecolor = blue, urlcolor =cyan]{hyperref}
\usepackage[figure]{hypcap}
\usepackage{xcolor}
\usepackage{mathtools}
\usepackage{mathdots}
\usepackage[notransparent]{svg}
\usepackage{todonotes} %I like to use this one to highlight missing bits for myself. -Alicia

\usepackage{array}
\newcolumntype{C}{>{$}c<{$}}
\AtBeginDocument{
\heavyrulewidth=.08em
\lightrulewidth=.05em
\cmidrulewidth=.03em
\belowrulesep=.65ex
\belowbottomsep=0pt
\aboverulesep=.4ex
\abovetopsep=0pt
\cmidrulesep=\doublerulesep
\cmidrulekern=.5em
\defaultaddspace=.5em
\tabcolsep=7pt
}
\usepackage{booktabs}
\usepackage{verbatim}

\usepackage{CJKutf8}

\makeatletter
\let\latex@addcontentsline\addcontentsline
\renewcommand{\addcontentsline}[3]{}
\makeatother

\definecolor{emerald}{rgb}{0.07, 0.53, 0.03}

\usepackage{braket,physics}

\usepackage{amsbsy}

\newcommand{\eq}[1]{Eq.~(\ref{eq:#1})}
\newcommand{\aLabel}{a}
\newcommand{\bLabel}{b}

\usepackage{nicefrac} %added for equations

\usepackage{soul}   % SOUL Added for underlines
\setstcolor{red} % strikethrough color
\setul{}{1.5pt}  % line thickness

\usepackage[dvipsnames]{xcolor}

\DeclareMathOperator{\diag}{diag}

\newcommand{\acd}{\tilde{\chi}_D}
\newcommand{\acm}{\tilde{\chi}_M}

\begin{document}

\title{In-situ Characterization of Light-Matter Coupling in Multimode Circuit-QED Systems}

\author{Kellen O'Brien}
\email{kobrie@umd.edu}
\affiliation{Department of Physics, University of Maryland, College Park, MD 20742, USA}
\affiliation{Joint Quantum Institute, NIST/University of Maryland, College Park, Maryland 20742 USA}
%\affiliation{Department of Physics and JQI, University of Maryland, College Park, MD 20742, USA}

\author{Won Chan Lee}
%\affiliation{Department of Physics and JQI, University of Maryland, College Park, MD 20742, USA}
\affiliation{Department of Physics, University of Maryland, College Park, MD 20742, USA}
\affiliation{Joint Quantum Institute, NIST/University of Maryland, College Park, Maryland 20742 USA}

\author{Alexandra Behne}
\affiliation{Joint Quantum Institute, NIST/University of Maryland, College Park, Maryland 20742 USA}
\affiliation{Joint Center for Quantum Information and Computer Science, NIST/University of Maryland, College Park, Maryland 20742 USA}

\author{Ali Fahimniya}
\affiliation{Joint Quantum Institute, NIST/University of Maryland, College Park, Maryland 20742 USA}
\affiliation{Joint Center for Quantum Information and Computer Science, NIST/University of Maryland, College Park, Maryland 20742 USA}

\author{\begin{CJK}{UTF8}{gbsn}Yu-Xin Wang (王语馨)\end{CJK}}
\affiliation{Joint Center for Quantum Information and Computer Science, NIST/University of Maryland, College Park, Maryland 20742 USA}

\author{Maya Amouzegar}
%\affiliation{Hopkins Applied Physics Lab}
\thanks{Current Address: Johns Hopkins Applied Physics Laboratory, Laurel, Maryland, 20723, USA.}
\affiliation{Department of Physics, University of Maryland, College Park, MD 20742, USA}
\affiliation{Joint Quantum Institute, NIST/University of Maryland, College Park, Maryland 20742 USA}

\author{Alexey V. Gorshkov}
\affiliation{Joint Quantum Institute, NIST/University of Maryland, College Park, Maryland 20742 USA}
\affiliation{Joint Center for Quantum Information and Computer Science, NIST/University of Maryland, College Park, Maryland 20742 USA}

\author{Alicia J.  Koll\'ar}
\affiliation{Department of Physics, University of Maryland, College Park, MD 20742, USA}
\affiliation{Joint Quantum Institute, NIST/University of Maryland, College Park, Maryland 20742 USA}
\affiliation{Maryland Quantum Materials Center, Department of Physics, University of Maryland, College Park, MD 20742, USA}
%\affiliation{Department of Physics, JQI, and QMC, University of Maryland, College Park, MD 20742, USA}

\preprint{APS/123-QED}

\date{March 5, 2026}% It is always \today, today,
             %  but any date may be explicitly specified

\begin{abstract}
Multimode cavity-QED systems can be leveraged to explore a wide range of physical phenomena; however, a complex multimode environment makes systematic characterization of light-matter interactions challenging.
Here we present a general measurement protocol, applicable to both atomic and synthetic cavity-QED systems, that enables the determination of coupling to individual photonic modes.
The method leverages measurements of the AC-Stark and Kerr effects, along with known detuning dependencies, to eliminate the need for single-photon resolution, independent photon-number calibration, or insertion-loss calibration. 
We demonstrate the method using a superconducting transmon qubit coupled to a one-dimensional microwave resonator lattice.
We validate the consistency of the extracted light-matter couplings $g$ determined at multiple qubit detunings, and from the self-Kerr and cross-Kerr shifts for three photon modes, which provide separate measurements of $g$ for each of the three modes.

\end{abstract}

%\keywords{Suggested keywords}%Use showkeys class option if keyword
                              %display desired
\maketitle

\noindent
Multimode cavity-QED systems can be leveraged to explore a wide range of physical phenomena, from phase transitions \cite{Kollar2017supermode, Gopalakrishnan10} to quantum spin models \cite{Lev:Ising,Douglas_2015,Gopalakrishnan:2011jx}.
The coupling between individual atom-like degrees of freedom and photon modes is fundamental to the dynamics of a realized system \cite{bienias:Hyperbolic};
however, these couplings can be challenging to determine in multimode systems,
due to the large variety of frequency splittings and coupling strengths present.

Here we present a measurement protocol that enables the determination of light-matter coupling to individual modes in multimode cavity-QED systems without the need for single-photon resolution~\cite{Schuster_2007:Number_Splitting,Wallraff_2004:VRS, Kimble_VRS, Miller_2005,Satzinger_2018}, independent photon-number calibration~\cite{Lee_PhotonCal,Kimble_mirrors,Weis_2010,Youssefi_2023}, or insertion-loss calibration~\cite{Kimble_mirrors,Thomas_cryo}. 
Instead, our method leverages the fundamental form of light-matter interaction,
using two different nonlinear effects: 
the AC-Stark shift \cite{steckquantum} on an off-resonant atom/qubit induced by driving one photon mode and the Kerr shifts~\cite{Haroche} on photon modes induced by the same drive.
Comparison of the power dependence of these two effects enables elimination of the unknown photon number.
The use of a second auxiliary photon mode makes it possible to determine the light-matter coupling of a chosen mode using only measurements on the auxiliary, which can be chosen independently of the mode of interest. 
Thus, the method presented here is applicable to all modes, even under-coupled or localized optical modes or phononic modes in hybrid phononic-cavity-QED systems~\cite{Satzinger_2018,Weis_2010,Youssefi_2023,quantumPhononicsReview,optomechanicsRevModPhys,Kippenberg2024} that are challenging to measure directly.

%Thus, the method presented here is applicable to all modes, even under-coupled or localized modes that are challenging to measure directly.
%In particular, this property makes the protocol well suited for determining coupling to phononic modes in optomechanical systems.

% The use of a second auxiliary photon mode makes it possible to determine the light-matter coupling of a chosen mode using only measurements on the auxiliary, which can be chosen independently of the mode of interest. 
% Thus, the method presented here is applicable to all modes, even under-coupled or localized modes that are challenging to measure directly.

We demonstrate this method using a superconducting transmon qubit~\cite{Blais:revmodphys} in a multimode coplanar waveguide (CPW) resonator lattice with $54$ modes, first described in Ref.~\cite{OBrien2025}.
We present a simple two-mode model describing a qubit coupled to the two chosen modes in a multimode device and derive equations for elimination of photon number dependence and extraction of the coupling strength $g_j$ to a chosen mode. 
We select three modes in the device and show that measurements for all possible pairs yield consistent values of $g_j$.
%Additionally, we tune the frequency of the transmon qubit over a range of $400$~MHz and show that the extracted values of $g_j$ are consistent across the full range.
Additionally, we vary the detuning between the transmon qubit and the photon modes from $400$~MHz to $800$~MHz and show that the extracted values of $g_j$ are consistent across the full range.
These results show that the simple method and model presented here yield accurate determination of the light-matter coupling strengths, even in highly multimode systems.

\begin{figure}[t]
\centering
    \includegraphics[width=0.45\textwidth]{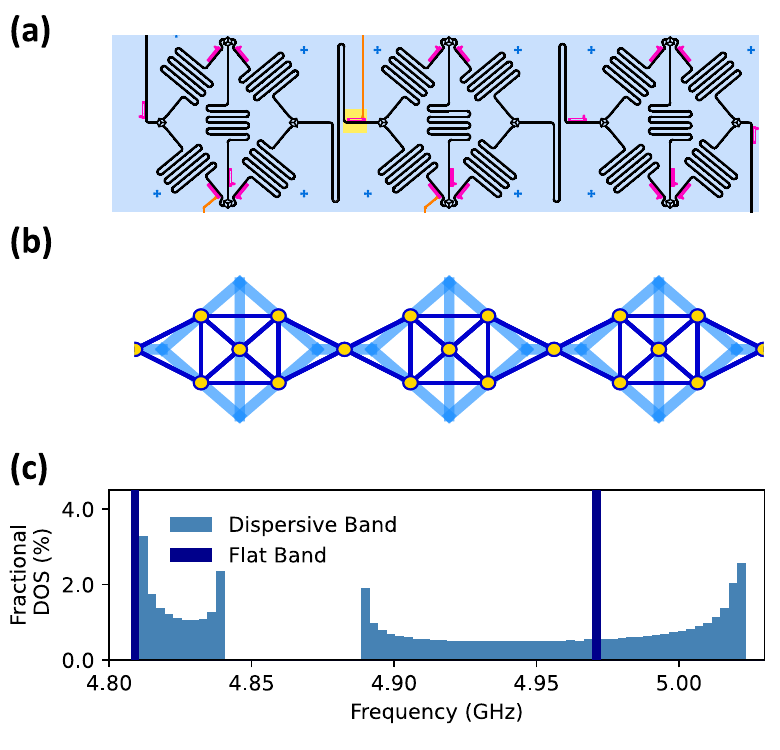}
	% \vspace{-0.6cm}
    \vspace{-0.2cm}
    \caption{\label{fig:device_image} 
    \textit{CPW lattice.} (a) CAD of three of the nine unit cells of the CPW lattice device used in this work. The resonators forming the lattice are shown in black, and the qubit used to perform the protocol is highlighted with a yellow square. 
    (Full device CAD is shown in Fig.~\ref{fig:mode_spectroscopy} in the Supplemental Material.) 
    (b) Schematic of the tight-binding lattice realized by this device overlaid on the physical layout of the resonator chain. Gold dots indicate lattice sites with dark blue lines indicating nearest-neighbor coupling. The light blue lines in the background represent the physical resonators. (c) Expected density of states for an infinite version of the quasi-1D tight-binding lattice displayed in (b). The dispersive bands are shown in light blue, and the flat bands, which extend above the scale of the plot, are shown in dark blue. The protocol is demonstrated using lattice modes between $4.96$ and $5.00$~GHz and qubit frequencies between $\sim 4.2$ and $\sim 4.6$~GHz.
    %\ko{new:} The modes utilized to demonstrate the protocol were selected from between $4.96$ and $5.00$~GHz.
    }
\end{figure}

% \section{CPW Lattice Device}
\vskip 0.05in
\noindent
\textit{CPW lattice device}.---Coplanar waveguide (CPW) resonator lattices~\cite{Koch:AnnPhysBerl,Houck_Nature_QS} with embedded transmon qubits are a platform which shows promise for the analog quantum simulation of spin-boson models.
The resonator array can be engineered to exhibit a wide range of exotic band structures and resonator connectivities, ranging from 1D to 2D to even non-Euclidean geometries \cite{Kollar:2019linegraph,Houck:earlylattice,Kollar:2019hyperbolic}, and
embedded transmon qubits couple strongly to the photonic bands and experience photon-mediated interactions which follow the connectivity of the resonator array~\cite{bienias:Hyperbolic, Douglas_2015, Sundaresan}.
Since the resulting photonic band structures contain large numbers of modes with different coupling strengths and degrees of localization~\cite{Kollar:2019linegraph, Kollar:2019hyperbolic}, methods for determining light-matter coupling strength which rely on single-photon resolved dispersive shifts~\cite{Schuster_2007:Number_Splitting} or vacuum Rabi splittings~\cite{Wallraff_2004:VRS} require very strong light-matter coupling or well-separated modes and are, therefore, challenging to apply systematically. This system is thus an ideal testbed for the protocol presented here, based on combined AC-Stark-shift and Kerr-shift measurements.

Before presenting the theory model in detail, we briefly describe the particular CPW lattice device upon which we perform our protocol~\cite{OBrien2025}.
The resonator array consists of nine unit cells with six resonators each; the resonators are coupled end-to-end to their nearest neighbors through three-way capacitors.
The computer-aided design (CAD) for three unit cells of the device is shown in Fig.~\ref{fig:device_image}(a).
The array is well-described by a bosonic tight-binding model, the sites of which correspond to resonators within the lattice. 
The on-site resonator frequency is primarily set by its length, and the hopping strength is set by the coupling capacitance \cite{Koch:AnnPhysBerl,Koch:TRS_breaking}.
In Fig.~\ref{fig:device_image}(b), the tight-binding lattice realized by this device is overlaid on a schematic of the resonator chain, and the density of states for an infinite version of the lattice is shown in Fig.~\ref{fig:device_image}(c).
The device contains three capacitively coupled flux-tunable transmons~\cite{OBrien2025}, only one of which [highlighted in Fig.~\ref{fig:device_image}(a)] is used in this work.
%Three flux-tunable transmons, one of which is highlighted in Fig.~\ref{fig:device_image}(a), are capacitively coupled to individual resonators within the bulk.

\vskip 0.05in
\noindent\textit{Background}.---In order to determine the coupling between the qubit and a chosen mode of the lattice, the protocol uses \emph{two} nondegenerate photonic modes and observations of responses to an external drive. Crucial to our extraction procedure are the AC-Stark shift on the qubit and the Kerr shifts on the photon modes induced by external drives: comparison of the power dependence of these two effects enables elimination of any unknown input coupling to the driven photon mode. To highlight the origins of those shifts,
we introduce the Hamiltonian describing a transmon qubit coupled to two modes in a CPW lattice, before discussing the measurement protocol in full.

For any chosen pair of modes, one will be referred to as the ``drive'' mode  ($D$), to which strong tones will be applied, and the other as the ``monitor'' mode ($M$), which will be used for detection. The total Hamiltonian for the qubit and these two modes is~\cite{Blais:revmodphys}
\begin{eqnarray}
    H_\text{tot} &=& H_0 + H_\text{int}, \label{eq:Htot_Hint}\\
    H_0 \ &=&  \omega_D d^{\dagger}d +  \omega_M m^{\dagger} m + \omega_q q^{\dagger} q-  \frac{\alpha}{2} q^{\dagger}q^{\dagger}qq, \label{eq:H0_twomode}\\
    H_\text{int} &=&  g_D(q^{\dagger} d + d^{\dagger} q) +  g_M(q^{\dagger} m + m^{\dagger} q). \label{eq:Hint_JC}
\end{eqnarray}
Here, $d^{\dagger}$ and $d$  ($m^{\dagger}$ and $m$) denote the bosonic raising and lowering operators of mode $D$ ($M$), $\omega_D$ ($\omega_M$) is the frequency of mode $D$ ($M$), and $g_D$ ($g_M$) is the coupling between the qubit and mode $D$ ($M$).
The transmon is effectively an anharmonic oscillator with frequency $\omega_q$, anharmonicity $\alpha$, and raising/lowering operators $q^{\dagger}$/$q$. 
While the transmon is treated as a two-level system in most contexts, the higher levels are explicitly included here as the magnitudes of the AC-Stark and Kerr effects depend on the anharmonicity.

In the dispersive limit where the qubit is far detuned from the photon modes, the interaction between the qubit and photon modes can be treated perturbatively. 
For low intracavity photon numbers, only second-order effects, such as the AC-Stark shift, are relevant.
In order to observe fourth-order Kerr effects, the system is intentionally driven to higher intracavity photon numbers.
Including these higher-order effects, the effective interaction Hamiltonian is given by 
\begin{equation}\label{eq:interactionH}
\begin{aligned}
H_{\text{int}} \approx & \   \acd d^{\dagger}d q^{\dagger}q +  \frac{\chi_{DD}}{2} d^{\dagger}d^{\dagger}dd  \\  &+  \acm m^{\dagger}m q^{\dagger}q +  \frac{\chi_{MM}}{2} m^{\dagger}m^{\dagger}mm \\ &+  \chi_{DM}d^{\dagger}dm^{\dagger}m.
\end{aligned}
\end{equation}
% \begin{equation}\label{eq:interactionH}
% \begin{aligned}
% H_{\text{int}} \approx &\hbar \chi^{(d)} d^{\dagger}d q^{\dagger}q + \hbar \frac{\chi_{\text{dd}}}{2} d^{\dagger}d^{\dagger}dd  \\  &+ \hbar \chi^{(m)}m^{\dagger}m q^{\dagger}q + \hbar \frac{\chi_{\text{mm}}}{2} m^{\dagger}m^{\dagger}mm \\ &+ \hbar \chi_{\text{dm}}d^{\dagger}dm^{\dagger}m.
% \end{aligned}
% \end{equation}
Here, $\acd$ ($\acm$) is the dispersive-coupling (AC-Stark) coefficient associated with mode $D$ ($M$), $\chi_{DD}$ ($\chi_{MM}$) is the self-Kerr coefficient for mode $D$ ($M$), and $\chi_{DM}$ is the cross-Kerr coefficient for modes $D$ and $M$.

The AC-Stark and Kerr coefficients are given by~\cite{zhang_drive-induced_2022,Elliott_2018,Blais:revmodphys}
\begin{align}\label{eq:chi_main}
\tilde{\chi}_j = \frac{2g_{j}^2 \alpha}{\Delta _{j} (\alpha - \Delta _{j})}, \ \ \ j = D, M;
\end{align}
\begin{equation}\label{eq:SelfKerrMain}
\chi_{jj} = \frac{2 g_j^4\alpha}{\Delta_j^3(\alpha - 2 \Delta_j)}, \ \ \ j = D,M;
\end{equation}
and
\begin{equation}\label{eq:CrossKerrMain}
\chi_{DM} = \frac{2 \alpha g_D^2 g_M^2(\Delta_D + \Delta_M)}{\Delta_D^2\Delta_M^2(\alpha-\Delta_D-\Delta_M)},
\end{equation}
where $\Delta_j \equiv \omega_q - \omega_j$.
%A general procedure for deriving these well-known coefficients from operator-valued Schrieffer-Wolff is described in the \ko{End Matter}, which further enables systematic calculation of dispersive shifts involving arbitrary transmon levels \ak{Theorists: write one sentence about why this is a useful new addition and modify the previous}.
% \ab{This procedure enables us to compute---in closed form as operators on the full transmon subspace---coefficients not only for the ground state of the transmon but also for each of the infinitely many excited states (when the transmon is approximated as an anharmonic oscillator).}
In the $|\alpha| \to \infty$ limit, above expressions simplify to well-known results for qubit-resonator coupling: $\tilde{\chi}_j \propto g_j^2/\Delta_j$, $\chi_{jj} \propto g_j^4/\Delta_j^3$, and $\chi_{DM} \propto g_D^2 g_M^2(\Delta_D +\Delta_M)/(\Delta_D^2 \Delta_M^2)$. 
In the End Matter we present a general procedure for deriving operator-valued versions of these well-known coefficients, which enables systematic calculation not only for a two level system or the ground state of a transmon, but also for arbitrary excited levels of the transmon.

\begin{figure*}[t]
\centering
    \includegraphics[width=1\textwidth]{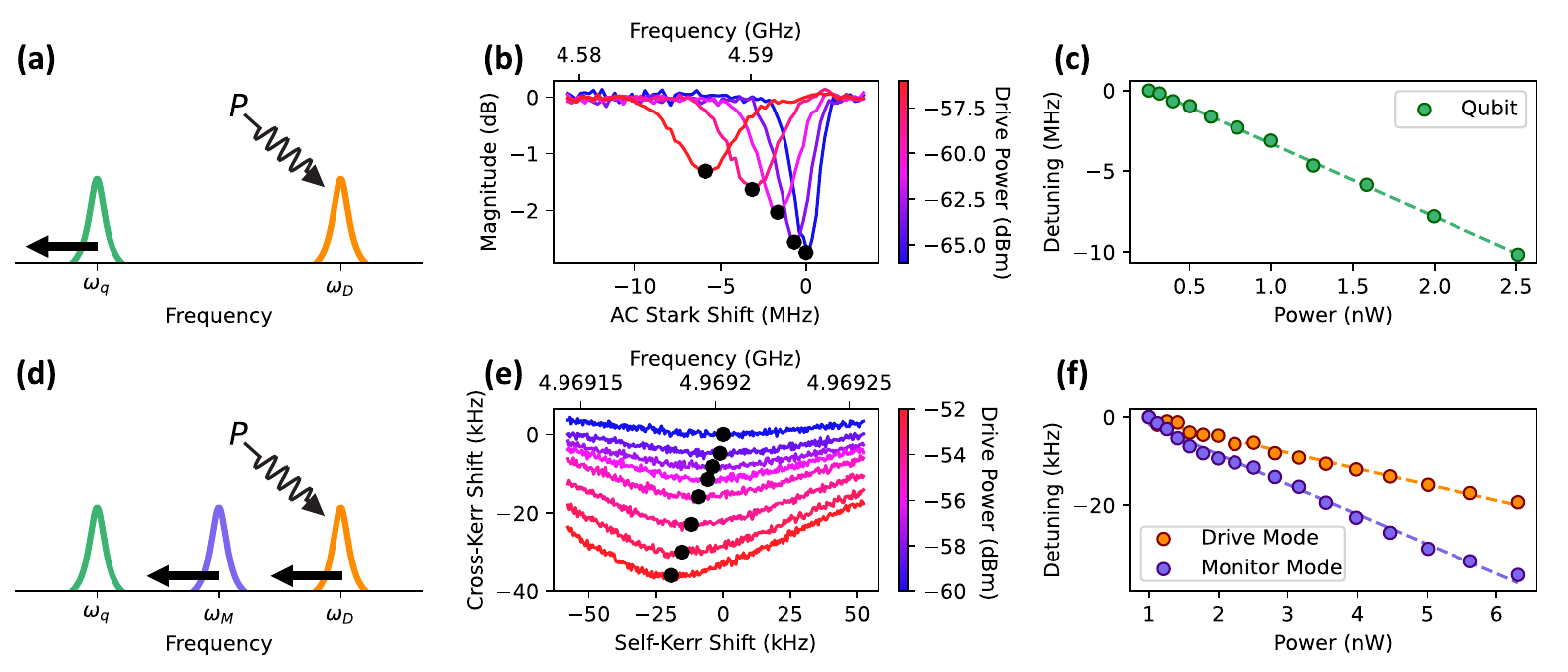}
	% \vspace{-0.6cm}
    \vspace{-0.2cm}
    \caption{\label{fig:measurement_protocol} \textit{AC-Stark and Kerr shifts.} Measurements of AC-Stark and Kerr shifts carried out using a drive mode with frequency $\omega_D/2\pi = 4.997$~GHz, a monitor mode with frequency $\omega_M/ 2\pi = 4.969$~GHz, and a qubit tuned to $\omega_q/2 \pi = 4.593$~GHz. (a) Schematic of the AC-Stark-shift measurement. Mode $D$ is driven while the AC-Stark shift of the qubit frequency, $\omega_q'$, is measured. 
    (b) Two-tone spectroscopy measurements~\cite{Blais:revmodphys} of the qubit with varying drive powers to mode $D$. (See Section~\ref{app:TwoTone} of the Supplemental Material for measurement details.)
    As the drive power increases, the effective qubit frequency (labeled with a black dot) is AC-Stark shifted down accordingly. (c) The measured qubit frequency shift versus the drive power. The slope of a linear fit to this data gives $\partial_{P} \omega_q'/2\pi= -32.2$~MHz/nW. (d) Schematic of the Kerr-shift measurement. Mode $D$ is driven while the self-Kerr and cross-Kerr shifts on $\omega_D'$ and $\omega_M'$, mediated by the qubit, are simultaneously measured. (e) Two-tone spectroscopy measurements of the drive mode $D$, taken by measuring transmission of a probe tone through mode $M$, with varying drive tone powers to mode $D$. As the drive power increases, the effective frequency of the drive mode (labeled with a black dot) is self-Kerr shifted down accordingly. The cross-Kerr shift on the monitor mode is inferred from the strength of the response at this point. (f) The measured frequency shift of the drive and monitor modes versus the drive power. A linear fit to the self-Kerr data yields a slope $\partial_{P} \omega_D'/2\pi = -32.0$~kHz/nW, and similarly for the cross-Kerr data, $\partial_{P} \omega_M'/2\pi = -42.1$~kHz/nW.}     
\end{figure*}

The coefficients in Eqs.~(\ref{eq:chi_main}-\ref{eq:CrossKerrMain}) are shifts of the qubit and normal-mode frequencies \emph{per photon}. 
Apart from the qubit-mode couplings, all three coefficients only involve transmon anharmonicity and detunings, which can be independently calibrated.
However, in order to extract the coefficients from experimental frequency shifts, we would also need precise knowledge of the intracavity photon numbers, which can be expensive to calibrate. In contrast, the shifts \emph{per drive power} are directly accessible in our experiments.

Here we show that the per-unit-power shifts can be used to extract both $g_D$ and $g_M$ without an independent photon number calibration.
To produce these frequency shifts, a resonant drive tone with variable power is applied to mode $D$, populating it with a coherent state.
As the drive power $P$ increases, the average intracavity photon number $\bar{n}_D \equiv \langle d^{\dagger} d \rangle$ in mode $D$ rises in tandem.
Following from Eq.~(\ref{eq:interactionH}), the increased photon number produces frequency shifts on both the qubit and resonator modes. The intracavity photon number is related to the applied drive power by a constant factor $\bar{n}_D = \beta P$. If $\beta$ is known exactly, then measurements of the AC-Stark-shifted qubit frequency can be directly converted to a measurement of $\acd$. However, the precise value of $\beta$ for a given drive mode is generally unknown, since it depends on many large systematic effects such as input attenuation and mode matching.

The AC-Stark shift produces a new effective qubit frequency $\omega_q' = \omega_q + \acd \bar{n}_D =  \omega_q + \acd \, \beta P$,
the self-Kerr shift an effective drive-mode frequency $\omega_D' = \omega_D + \chi_{DD} \bar{n}_D = \omega_D + \chi_{DD} \, \beta P$, and the cross-Kerr shift an effective monitor mode frequency $\omega_M' = \omega_M + \chi_{DM} \bar{n}_D = \omega_M + \chi_{DM} \, \beta P$.
Contributions from the fixed photon number in the monitor mode are assumed to either be negligible or be absorbed into the bare mode frequencies.
% , and the constant dispersive shift from the ground-state of the qubit is neglected from the effective monitor-mode and drive-mode frequencies for simplicity.
%Contributions from the fixed photon number in the monitor mode or the dispersive shift of the ground-state qubit are constant and therefore have been absorbed into the bare mode frequencies for simplicity.
The frequency shifts per unit drive power $\partial_P \omega' = \partial \omega' / \partial {P}$ can readily be determined from measurements at different drive powers, and taking ratios of these removes all $\beta$ dependence. From Eqs.~(\ref{eq:chi_main}--\ref{eq:CrossKerrMain}) we find
\begin{equation}\label{eq:SelfKerrRatio}
\frac{\partial_{P} \omega_D'}{\partial_{P} \omega_q'}   = 
\frac{\beta \, \chi_{DD}}{\beta \, \acd} = 
\frac{-g_D^2 (\alpha - \Delta_D)}{\Delta_D^2 (\alpha - 2\Delta_D)}
\end{equation}
and
\begin{equation}\label{eq:CrossKerrRatio}
\frac{\partial_{P} \omega_M'}{\partial_{P} \omega_q'}    =  
\frac{\beta \, \chi_{DM}}{\beta \, \acd} = \frac{-g_M^2 (\Delta_D + \Delta_M)(\alpha - \Delta_D)}{\Delta_D\Delta_M^2(\alpha-\Delta_D-\Delta_M)}.
\end{equation}
As the transmon anharmonicity and detunings can be determined from standard spectroscopy measurements, the qubit-mode couplings $g_D$ and $g_M$ can be extracted directly from Eqs.~\eqref{eq:SelfKerrRatio} and~\eqref{eq:CrossKerrRatio}.

\noindent\textit{Measurement protocol}.---To determine the AC-Stark-shifted qubit frequency, we use a pump-probe spectroscopy method~\cite{Schoelkopf:Two_Tone} based on dispersive readout of superconducting qubits~\cite{Blais:revmodphys}. 
The nonlinearity of the qubit combined with strong light-matter coupling produces a non-negligible qubit-state-dependent frequency shift $\tilde{\chi}_j$ on a host resonator. As a result, a microwave tone probing the photonic mode experiences amplitude and phase shifts depending on the qubit state.
Sweeping a pump tone and measuring the nonlinear response in the probe when the pump comes into resonance with the qubit, a measurement scheme known as two-tone spectroscopy \cite{Schoelkopf:Two_Tone}, allows the qubit frequency to be identified spectroscopically using direct measurements of the resonator only.

While two-tone spectroscopy is normally used to measure a highly anharmonic qubit with a single strongly coupled resonator, the same mechanism translates readily to determination of the self-Kerr and cross-Kerr shifts between two weakly anharmonic modes~\cite{Elliott_2018,Kirchmair_2013}.
As in the qubit case, the effective frequency of one mode in the presence of the self-Kerr shift from the drive can be identified as the pump frequency which maximizes the induced frequency shifts on the second \cite{Peugeot_2024,Bosman_2017}.
In particular, this allows modes with very low transmission, e.g. localized flat-band modes, to be characterized via their coupling to easy-to-measure modes with strong transmission. 
Simultaneously, the cross-Kerr shift on the second mode can be extracted from the size of the amplitude and phase shifts on the probe tone (see Section~\ref{app:TwoTone} in the Supplemental Material).

To obtain the partial derivatives in Eqs.~(\ref{eq:SelfKerrRatio}--\ref{eq:CrossKerrRatio}), we apply a drive tone to mode $D$ with variable power $P$ and carry out two types of measurements.
In case 1, %we apply a drive 
the drive is applied to mode $D$ to AC-Stark shift the qubit frequency, as shown schematically in Fig.~\ref{fig:measurement_protocol}(a).
In case 2, %a stronger drive 
the drive is stronger and is applied to mode $D$ to produce Kerr shifts on both mode $D$ and mode $M$, as shown schematically in Fig.~\ref{fig:measurement_protocol}(d).
In case 1, we use two-tone spectroscopy with a pump on the qubit to measure $\omega_q'$ versus $P$ and determine $\partial_P \omega_q'$ (see Section~\ref{app:Protocol_Details} of the Supplemental Material for details).
Sample data for this process are shown in Fig.~\ref{fig:measurement_protocol}(b) and (c).
At these intermediate values of $P$, the second-order AC-Stark shift is significant, while the fourth-order self-Kerr shift on mode $D$ is negligible. 
In case 2, the strong drive applied to mode $D$ produces a self-Kerr shift in mode $D$ and a cross-Kerr shift in mode $M$ that are both comparable in size. To ensure that 
$\beta$ is constant 
over a wide range of drive powers, the drive is scanned over mode $D$ at each power to locate the modified resonance frequency $\omega_D'$ in the presence of the self-Kerr shift (see Section~\ref{app:semiclassic} of the Supplemental Material), identified as the drive frequency which produces the largest cross-Kerr shift. 
$\omega_M'$ at this resonant drive frequency is obtained simultaneously.
Sample data are shown in Fig.~\ref{fig:measurement_protocol}(e) and (f).
See Section~\ref{app:Protocol_Details} of the Supplemental Material for details of the determination of $\omega_D'$ and $\omega_M'$ and the conversion from monitor tone transmission to frequency shifts.

\vskip 0.05in
\noindent \textit{Protocol demonstration}.--- 
The model underlying the measurement protocol introduced above includes only a transmon coupled to two modes.
However, the full CPW lattice device hosts dozens of photonic modes with frequencies that lie between $4.8$ and $5.1$~GHz.
To verify that the qubit-mode couplings extracted from the model are consistent despite the multimode environment, we perform the measurement protocol on three lattice modes using all six possible combinations of drive and monitor mode pairings.
Figure~\ref{fig:measurement_protocol} shows the data and output of the protocol for one such pair using a drive mode with frequency $\omega_D/2\pi = 4.969$~GHz, a monitor mode with frequency $\omega_M/2\pi = 4.960$~GHz, and a transmon qubit tuned to the frequency $\omega_q/2\pi = 4.593$~GHz with anharmonicity $\alpha/2\pi = 113$~MHz.
In this configuration, the qubit is detuned by more than $200$~MHz from the lowest-frequency lattice mode [see Fig.~\ref{fig:device_image}(c) for reference].
The results for the remaining five pairs can be found in Section~\ref{app:3Modes} of the Supplemental Material.

To observe the AC-Stark shift, two-tone spectroscopy measurements of the qubit were collected at nine drive powers, a subset of which are displayed in Fig.~\ref{fig:measurement_protocol}(b).
The AC-Stark shift per unit power for the chosen qubit and drive mode, determined by a linear fit to the extracted qubit frequencies shown in Fig.~\ref{fig:measurement_protocol}(c), is found to be $\partial_{P} \omega_q'/2\pi = -4.52 \pm 0.05$~MHz/nW.
Two-tone spectroscopy measurements that simultaneously track the self-Kerr and cross-Kerr-shifted frequencies of modes $D$ and $M$ are collected at 17 drive powers, a subset of which are displayed in Fig.~\ref{fig:measurement_protocol}(e).
The self-Kerr shift per unit power and the cross-Kerr shift per unit power, determined by linear fits to the extracted mode frequencies shown in Fig.~\ref{fig:measurement_protocol}(f), are found to be $\partial_{P} \omega_D'/2 \pi = -3.62 \pm 0.09$~kHz/nW and $\partial_{P} \omega_M'/2 \pi = -6.8 \pm 0.2$~kHz/nW.
%Taking the ratios between $\beta \chi_{DD}$ and $\beta \chi_{DM}$ with $\beta \chi^{(D)}$ and using the relations 
From Eqs.~\eqref{eq:SelfKerrRatio} and \eqref{eq:CrossKerrRatio}, the qubit-mode couplings are found to be $g_D = 14.2 \pm 0.6$~MHz for the drive mode and $g_M = 13.4 \pm 0.5$~MHz for the monitor mode.
Error bars for $g$ were determined from repeatability measurements, where systematic variation, likely due to qubit drift, was found to dominate over statistical error (see Section~\ref{app:stability} of the Supplemental Material for details).
%\ko{Explain sources of error bars, first time for the Stark/Kerr slopes and bring up the g slope so it occurs earlier}

\begin{figure*}[t]
\centering
    \includegraphics[width=1\textwidth]{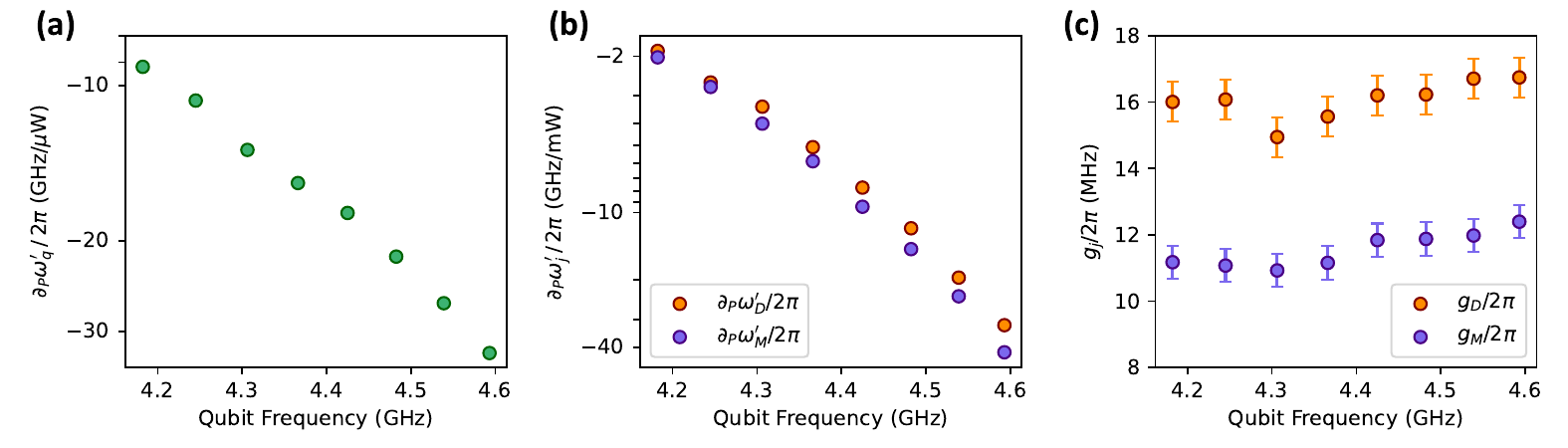}
	% \vspace{-0.6cm}
    \vspace{-0.2cm}
    \caption{\label{fig:detuning_sweep} 
    \textit{Consistency of qubit-mode coupling versus detuning.}
    Results from repeated iterations of the $g$-verification protocol using a drive mode with frequency $\omega_D/2\pi = 4.997$~GHz, a monitor mode with frequency $\omega_M/2\pi = 4.960$~GHz, and with the qubit frequency swept from $\sim 4.2$~GHz -- $\sim 4.6$~GHz. 
    The slopes (a) $\partial_{P} \omega_q'$ and (b) $\partial_{P} \omega_D'$ and $\partial_{P} \omega_M'$ at each detuning, determined from AC-Stark-shift and Kerr-shift measurements, respectively. Fit error bars are smaller than markers and not shown.
    (c) Calculated qubit-mode couplings for the drive mode ($g_D$) and the monitor mode ($g_M$). 
    Error bars indicate expected systematic error determined from repeated measurement at two fixed detunings (see Section~\ref{app:stability} of the Supplemental Material). The calculated couplings at different detunings are consistent with a drive-mode coupling $g_D/2\pi$ of approximately $ 16$~MHz and a monitor-mode coupling $g_M/2\pi$ of approximately $12$~MHz.} 
  
\end{figure*}

To verify that the analysis accurately accounts for the frequency dependence of the transmon [see Eqs.~\eqref{eq:SelfKerrRatio} and \eqref{eq:CrossKerrRatio}] in the presence of more than two modes, 
we repeat the protocol with the qubit tuned through a range of frequencies and verify that the estimates for $g_D$ and $g_M$ are consistent. These results, shown in Fig.~\ref{fig:detuning_sweep}, utilize a drive mode with frequency $\omega_D/2\pi = 4.997$~GHz and a monitor mode with frequency $\omega_M/2\pi = 4.960$~GHz.
The qubit was tuned to $8$ frequencies between $\sim 4.2$ and $\sim 4.6$~GHz, and the frequency shifts per unit power were extracted from the AC-Stark and Kerr shift measurements at each detuning, displayed in Fig.~\ref{fig:detuning_sweep}(a) and Fig.~\ref{fig:detuning_sweep}(b), respectively.

Despite the AC-Stark and Kerr shifts per unit power varying drastically with qubit frequency, as expected from Eqs.~(\ref{eq:chi_main}--\ref{eq:CrossKerrMain}), the calculated couplings at different detunings, shown in Fig.~\ref{fig:detuning_sweep}(c), are consistent with a drive-mode coupling $g_D/2\pi$ of approximately $ 16$~MHz and a monitor-mode coupling $g_M/2\pi$ of approximately $12$~MHz.
While $g$ in general is proportional to $\sqrt{\omega_q'}$ \cite{Koch:transmon}, for the relatively small range of qubit frequencies in the sample the coupling is expected to remain mostly stable, only varying by $5 \%$ between the highest and lowest frequencies.

% \noindent \textit{Discussion}.---We have introduced a measurement protocol for determining qubit-mode couplings and demonstrated it using a transmon qubit coupled to the photonic modes of a superconducting CPW resonator lattice. 
% The method, which uses measurements of AC-Stark and Kerr effects to calibrate out an unknown drive power to photon number conversion factor, was demonstrated for a range of qubit frequencies and using different mode combinations to ensure the robustness of the predicted coupling.
% Of particular note, this protocol can be used to determine couplings to modes that are not observable in transmission, such as localized flat-band modes or photonic bound states.
% The development of these techniques expands the measurement toolkit of multimode cavity-QED, allowing for the characterization of light-matter interactions in a wide range of multimode systems.

%%%%%%%%%
\vskip 0.05in
\noindent \textit{Discussion}.--- The protocol developed here expands the measurement toolkit of multimode cavity-QED and allows for characterization of interaction strengths in a wide range of multimode systems. It was demonstrated here using a transmon qubit coupled to microwave CPW resonators, but generalizes readily to any atom-like object coupled to \emph{bosonic} modes, simply by using the corresponding version of the AC-Stark and Kerr coefficients in Eqs.~(\ref{eq:chi_main}--\ref{eq:CrossKerrMain}), which can be calculated using the general operator-valued Schrieffer-Wolf procedure presented in the End Matter. 
Furthermore, a single, easily measured, auxiliary monitor mode enables characterization of coupling to other modes which are challenging to measure directly. For example, delocalized modes with high transmission can be used to extract couplings to localized/undercoupled modes in flat bands or moir\'e systems, and optical modes in hybrid phononic-cavity-QED systems can be used to determine couplings to phonon modes~\cite{Satzinger_2018,Weis_2010,Youssefi_2023,quantumPhononicsReview,optomechanicsRevModPhys,Kippenberg2024}, for which direct detector technology does not exist.

\vskip 0.1in
\textbf{Acknowledgments:} This research was supported by the Air Force Office of Scientific Research (Grant No. FA9550-21-1-0129),the National Science Foundation (QLCI grant OMA-2120757, PHY2047732, and PFC at JQI PHY-1430094), the Sloan Foundation, and the Maryland Quantum Materials Center. A.B., A.F., and A.V.G.~were supported in part by ARL (W911NF-24-2-0107), NSF QLCI (award No.~OMA-2120757), the DoE ASCR Quantum Testbed Pathfinder program (award No.~DE-SC0024220), ONR MURI,  NSF STAQ program, AFOSR MURI,  DARPA SAVaNT ADVENT,  and NQVL:QSTD:Pilot:FTL. A.B., A.F., and A.V.G.~also acknowledge support from the U.S.~Department of Energy, Office of Science, National Quantum Information Science Research Centers, Quantum Systems Accelerator (award No.~DE-SCL0000121) and from the U.S.~Department of Energy, Office of Science, Accelerated Research in Quantum Computing, Fundamental Algorithmic Research toward Quantum Utility (FAR-Qu). Y.-X.W.~acknowledges support from a QuICS Hartree Postdoctoral Fellowship. 
% \end{acknowledgements}

%%%%%%%%%%%%%%%%%
%\ko{sidetracked note: crit photon number estimated to be 250 photons. Max Stark shift photon number is ~100 photons, max Kerr shift photon number is ~250 photons.}
%\appendix

%\section{Protocol Details}

%\section{Measurement Uncertainty}

%\section{Critical Photon Number idk}

%$\omega_q$ 
%Description: Dressed qubit frequency.
%Method: Extract qubit frequency from the center frequency of Lorentzian fit. We'll use the lowest cavity power trace of the scan so as to minimize Stark shift's effect on final frequency. 
%$\Delta \omega_q$: Fit will return the error in the fitted center frequency. 

%Parameters: 
%$\alpha$ : We extract this from an old high power spec data set that has both the $\vert 0 \rangle \rightarrow \vert 1\rangle$ transition and $(\vert 0 \rangle \rightarrow \vert 2\rangle)/2$ transition. $\Delta \alpha$: We'll go back, do a Lorentzian fit to get the center frequencies of each transition, and propagate this out. 

%%\bibliographystyle{alpha}
\bibliographystyle{apsrev4-2}
\bibliography{refs.bib}

@Article{Kollar2017supermode,
author={Koll{\'a}r, Alicia J.
and Papageorge, Alexander T.
and Vaidya, Varun D.
and Guo, Yudan
and Keeling, Jonathan
and Lev, Benjamin L.},
title={Supermode-density-wave-polariton condensation with a Bose--Einstein condensate in a multimode cavity},
journal={Nat. Comm.},
year={2017},
month={Feb},
day={17},
volume={8},
number={1},
pages={14386},
issn={2041-1723},
doi={10.1038/ncomms14386},
url={https://doi.org/10.1038/ncomms14386}
}

@article{Kollar:2019linegraph,
   title={Line-Graph Lattices: Euclidean and Non-Euclidean Flat Bands, and Implementations in Circuit Quantum Electrodynamics},
   volume={376},
   ISSN={1432-0916},
   url={http://dx.doi.org/10.1007/s00220-019-03645-8},
   DOI={10.1007/s00220-019-03645-8},
   number={3},
   journal={Commun. Math. Phys.},
   publisher={Springer Science and Business Media LLC},
   author={Koll\'{a}r, Alicia J. and Fitzpatrick, Mattias and Sarnak, Peter and Houck, Andrew A.},
   year={2019},
   month=dec, pages={1909–1956} }

@Article{Kollar:2019hyperbolic,
author={Koll{\'a}r, Alicia J.
and Fitzpatrick, Mattias
and Houck, Andrew A.},
title={Hyperbolic lattices in circuit quantum electrodynamics},
journal={Nature},
year={2019},
month={Jul},
day={01},
volume={571},
number={7763},
pages={45-50},
issn={1476-4687},
doi={10.1038/s41586-019-1348-3},
url={https://doi.org/10.1038/s41586-019-1348-3}
}

@article{Sundaresan,
	title = {Interacting Qubit-Photon Bound States with Superconducting Circuits},
	author = {Sundaresan, Neereja M. and Lundgren, Rex and Zhu, Guanyu and Gorshkov, Alexey V. and Houck, Andrew A.},
	journal = {Phys. Rev. X},
	volume = {9},
	issue = {1},
	pages = {011--021},
	numpages = {23},
	year = {2019},
	month = {Feb},
	publisher = {American Physical Society},
	doi = {10.1103/PhysRevX.9.011021},
	url = {https://link.aps.org/doi/10.1103/PhysRevX.9.011021}
}

@article{Koch:transmon,
  title = {Charge-insensitive qubit design derived from the Cooper pair box},
  author = {Koch, Jens and Yu, Terri M. and Gambetta, Jay and Houck, A. A. and Schuster, D. I. and Majer, J. and Blais, Alexandre and Devoret, M. H. and Girvin, S. M. and Schoelkopf, R. J.},
  journal = {Phys. Rev. A},
  volume = {76},
  issue = {4},
  pages = {042319},
  numpages = {19},
  year = {2007},
  month = {Oct},
  publisher = {American Physical Society},
  doi = {10.1103/PhysRevA.76.042319},
  url = {https://link.aps.org/doi/10.1103/PhysRevA.76.042319}
}

@article{Peugeot_2024,
   title={Two-tone spectroscopy of high-frequency quantum circuits with a Josephson emitter},
   volume={22},
   ISSN={2331-7019},
   url={http://dx.doi.org/10.1103/PhysRevApplied.22.064027},
   DOI={10.1103/physrevapplied.22.064027},
   number={6},
   journal={Physical Review Applied},
   publisher={American Physical Society (APS)},
   author={Peugeot, A. and Riechert, H. and Annabi, S. and Balembois, L. and Villiers, M. and Flurin, E. and Griesmar, J. and Arrighi, E. and Pillet, J.-D. and Bretheau, L.},
   year={2024},
   month=dec }

@article{Bosman_2017,
   title={Multi-mode ultra-strong coupling in circuit quantum electrodynamics},
   volume={3},
   ISSN={2056-6387},
   url={http://dx.doi.org/10.1038/s41534-017-0046-y},
   DOI={10.1038/s41534-017-0046-y},
   number={1},
   journal={npj Quantum Information},
   publisher={Springer Science and Business Media LLC},
   author={Bosman, Sal J. and Gely, Mario F. and Singh, Vibhor and Bruno, Alessandro and Bothner, Daniel and Steele, Gary A.},
   year={2017},
   month=oct }

@article{Wallraff_2004:VRS,
   title={Strong coupling of a single photon to a superconducting qubit using circuit quantum electrodynamics},
   volume={431},
   ISSN={1476-4687},
   url={http://dx.doi.org/10.1038/nature02851},
   DOI={10.1038/nature02851},
   number={7005},
   journal={Nature},
   publisher={Springer Science and Business Media LLC},
   author={Wallraff, A. and Schuster, D. I. and Blais, A. and Frunzio, L. and Huang, R.- S. and Majer, J. and Kumar, S. and Girvin, S. M. and Schoelkopf, R. J.},
   year={2004},
   month=sep, pages={162–167} }

@article{Koch:TRS_breaking,
  title = {Time-reversal-symmetry breaking in circuit-QED-based photon lattices},
  author = {Koch, Jens and Houck, Andrew A. and Hur, Karyn Le and Girvin, S. M.},
  journal = {Phys. Rev. A},
  volume = {82},
  issue = {4},
  pages = {043811},
  numpages = {18},
  year = {2010},
  month = {Oct},
  publisher = {American Physical Society},
  doi = {10.1103/PhysRevA.82.043811},
  url = {https://link.aps.org/doi/10.1103/PhysRevA.82.043811}
}

@article{Koch:AnnPhysBerl,
   title={Circuit QED lattices: Towards quantum simulation with superconducting circuits},
   volume={525},
   ISSN={1521-3889},
   url={http://dx.doi.org/10.1002/andp.201200261},
   DOI={10.1002/andp.201200261},
   number={6},
   journal={Ann. Phys.},
   publisher={Wiley},
   author={Schmidt, Sebastian and Koch, Jens},
   year={2013},
   month=apr, pages={395–412} }

@article{Blais:revmodphys,
  title = {Circuit quantum electrodynamics},
  author = {Blais, Alexandre and Grimsmo, Arne L. and Girvin, S. M. and Wallraff, Andreas},
  journal = {Rev. Mod. Phys.},
  volume = {93},
  issue = {2},
  pages = {025005},
  numpages = {72},
  year = {2021},
  month = {May},
  publisher = {American Physical Society},
  doi = {10.1103/RevModPhys.93.025005},
  url = {https://link.aps.org/doi/10.1103/RevModPhys.93.025005}
}

@article{Houck_Nature_QS,
	author = {Andrew A. Houck and Hakan E. T{\"u}reci and Jens Koch},
	doi = {10.1038/nphys2251},
	issn = {17452473},
	journal = {Nat. Phys.},
	language = {eng},
	pages = {292{\textendash}299},
	publisher = {Nature Publishing Group},
	title = {On-chip quantum simulation with superconducting circuits},
	url = {http://dx.doi.org/10.1038/nphys2251},
	volume = {8},
	year = {2012}}

@article{Douglas_2015,
   title={Quantum many-body models with cold atoms coupled to photonic crystals},
   volume={9},
   ISSN={1749-4893},
   url={http://dx.doi.org/10.1038/nphoton.2015.57},
   DOI={10.1038/nphoton.2015.57},
   number={5},
   journal={Nat. Phot.},
   publisher={Springer Science and Business Media LLC},
   author={Douglas, J. S. and Habibian, H. and Hung, C.-L. and Gorshkov, A. V. and Kimble, H. J. and Chang, D. E.},
   year={2015},
   month=apr, pages={326–331} }

@article{Houck:earlylattice,
  title = {Low-disorder microwave cavity lattices for quantum simulation with photons},
  author = {Underwood, D. L. and Shanks, W. E. and Koch, Jens and Houck, A. A.},
  journal = {Phys. Rev. A},
  volume = {86},
  issue = {2},
  pages = {023837},
  numpages = {5},
  year = {2012},
  month = {Aug},
  publisher = {American Physical Society},
  doi = {10.1103/PhysRevA.86.023837},
  url = {https://link.aps.org/doi/10.1103/PhysRevA.86.023837}
}

@article{Gopalakrishnan10, 
year = {2010}, 
title = {{Atom-light crystallization of Bose-Einstein condensates in multimode cavities: Nonequilibrium classical and quantum phase transitions, emergent lattices, supersolidity, and frustration}}, 
author = {Gopalakrishnan, Sarang and Lev, Benjamin L. and Goldbart, Paul M}, 
journal = {Phys. Rev. A}, 
url = {https://doi.org/10.1103/PhysRevA.82.043612}, 
number = {4}, 
volume = {82}, 
language = {English}
}

@article{Gopalakrishnan:2011jx, 
year = {2011}, 
title = {{Frustration and Glassiness in Spin Models with Cavity-Mediated Interactions}}, 
author = {Gopalakrishnan, Sarang and Lev, Benjamin L. and Goldbart, Paul M}, 
journal = {Phys. Rev. Lett.}, 
doi = {10.1103/physrevlett.107.277201}, 
pages = {277201}, 
number = {27}, 
volume = {107}, 
language = {English}
}

@article{bienias:Hyperbolic, 
year = {2022}, 
title = {{Circuit Quantum Electrodynamics in Hyperbolic Space: From Photon Bound States to Frustrated Spin Models}}, 
author = {Bienias, Przemyslaw and Boettcher, Igor and Belyansky, Ron and Koll\'{a}r, Alicia J. and Gorshkov, Alexey V.}, 
journal = {Phys. Rev. Lett.}, 
issn = {0031-9007}, 
doi = {10.1103/physrevlett.128.013601}, 
pmid = {35061450}, 
pages = {013601}, 
number = {1}, 
volume = {128}, 
}

@article{Elliott_2018,
	author = {Elliott, Matthew and Joo, Jaewoo and Ginossar, Eran},
	doi = {10.1088/1367-2630/aa9243},
	journal = {New J. Phys.},
	month = {feb},
	number = {2},
	pages = {023037},
	publisher = {IOP Publishing},
	title = {Designing Kerr interactions using multiple superconducting qubit types in a single circuit},
	url = {https://dx.doi.org/10.1088/1367-2630/aa9243},
	volume = {20},
	year = {2018}}

@article{Schuster_2007:Number_Splitting,
   title={Resolving photon number states in a superconducting circuit},
   volume={445},
   ISSN={1476-4687},
   url={http://dx.doi.org/10.1038/nature05461},
   DOI={10.1038/nature05461},
   number={7127},
   journal={Nature},
   publisher={Springer Science and Business Media LLC},
   author={Schuster, D. I. and Houck, A. A. and Schreier, J. A. and Wallraff, A. and Gambetta, J. M. and Blais, A. and Frunzio, L. and Majer, J. and Johnson, B. and Devoret, M. H. and Girvin, S. M. and Schoelkopf, R. J.},
   year={2007},
   month=feb, pages={515–518} }

@article{OBrien2025,
  title = {Circuit-QED Lattice System with Flexible Connectivity and Gapped Flat Bands for Photon-Mediated Spin Models},
  author = {O'Brien, Kellen and Amouzegar, Maya and Lee, Won Chan and Ritter, Martin and Koll\'ar, Alicia J.},
  journal = {PRX Quantum},
  volume = {7},
  issue = {1},
  pages = {010321},
  numpages = {28},
  year = {2026},
  month = {Jan},
  publisher = {American Physical Society},
  doi = {10.1103/z9n2-mmmf},
  url = {https://link.aps.org/doi/10.1103/z9n2-mmmf}
}

@article{zhang_drive-induced_2022,
	title = {Drive-induced nonlinearities of cavity modes coupled to a transmon ancilla},
	volume = {105},
	issn = {2469-9926, 2469-9934},
	url = {https://link.aps.org/doi/10.1103/PhysRevA.105.022423},
	doi = {10.1103/PhysRevA.105.022423},
	pages = {022423},
	number = {2},
	journal = {Phys. Rev. A},
year = {2022},
month = {Feb},
	author = {Zhang, Yaxing and Curtis, Jacob C. and Wang, Christopher S. and Schoelkopf, R. J. and Girvin, S. M.},
}

@article{Walls1980,
doi = {10.1088/0305-4470/13/2/034},
url = {https://dx.doi.org/10.1088/0305-4470/13/2/034},
year = {1980},
month = {feb},
publisher = {},
volume = {13},
number = {2},
pages = {725},
author = {P D Drummond and  D F Walls},
title = {Quantum theory of optical bistability. I. Nonlinear polarisability model},
journal = {Journal of Physics A: Mathematical and General}
}

@article{Ciuti2016,
  title = {Exact steady state of a Kerr resonator with one- and two-photon driving and dissipation: Controllable Wigner-function multimodality and dissipative phase transitions},
  author = {Bartolo, Nicola and Minganti, Fabrizio and Casteels, Wim and Ciuti, Cristiano},
  journal = {Phys. Rev. A},
  volume = {94},
  issue = {3},
  pages = {033841},
  numpages = {11},
  year = {2016},
  month = {Sep},
  publisher = {American Physical Society},
  doi = {10.1103/PhysRevA.94.033841},
  url = {https://link.aps.org/doi/10.1103/PhysRevA.94.033841}
}

@article{Clerk2020,
  title = {Driven-Dissipative Quantum Kerr Resonators: New Exact Solutions, Photon Blockade and Quantum Bistability},
  author = {Roberts, David and Clerk, Aashish A.},
  journal = {Phys. Rev. X},
  volume = {10},
  issue = {2},
  pages = {021022},
  numpages = {27},
  year = {2020},
  month = {Apr},
  publisher = {American Physical Society},
  doi = {10.1103/PhysRevX.10.021022},
  url = {https://link.aps.org/doi/10.1103/PhysRevX.10.021022}
}

@book{steckquantum,
  title     = "Quantum and Atom Optics",
  author    = "Daniel A. Steck",
  year      = 2007
}

@article{Kippenberg2024, title={Room-temperature quantum optomechanics using an ultralow noise cavity}, volume={626}, url={http://dx.doi.org/10.1038/s41586-023-06997-3}, DOI={10.1038/s41586-023-06997-3}, number={7999}, journal={Nature}, publisher={Springer Science and Business Media LLC}, author={Huang, Guanhao and Beccari, Alberto and Engelsen, Nils J. and Kippenberg, Tobias J.}, year={2024}, month=feb, pages={512–516} }

@article{Schoelkopf:Two_Tone,
  title = {ac Stark Shift and Dephasing of a Superconducting Qubit Strongly Coupled to a Cavity Field},
  author = {Schuster, D. I. and Wallraff, A. and Blais, A. and Frunzio, L. and Huang, R.-S. and Majer, J. and Girvin, S. M. and Schoelkopf, R. J.},
  journal = {Phys. Rev. Lett.},
  volume = {94},
  issue = {12},
  pages = {123602},
  numpages = {4},
  year = {2005},
  month = {Mar},
  publisher = {American Physical Society},
  doi = {10.1103/PhysRevLett.94.123602},
  url = {https://link.aps.org/doi/10.1103/PhysRevLett.94.123602}
}

@article{Lev:Ising,
  title = {Multimode Cavity QED Ising Spin Glass},
  author = {Marsh, Brendan P. and Schuller, David Atri and Ji, Yunpeng and Hunt, Henry S. and Socolof, Giulia Z. and Bowman, Deven P. and Keeling, Jonathan and Lev, Benjamin L.},
  journal = {Phys. Rev. Lett.},
  volume = {135},
  issue = {16},
  pages = {160403},
  numpages = {7},
  year = {2025},
  month = {Oct},
  publisher = {American Physical Society},
  doi = {10.1103/x19r-pzyb},
  url = {https://link.aps.org/doi/10.1103/x19r-pzyb}
}

@article{Kirchmair_2013,
   title={Observation of quantum state collapse and revival due to the single-photon Kerr effect},
   volume={495},
   ISSN={1476-4687},
   url={http://dx.doi.org/10.1038/nature11902},
   DOI={10.1038/nature11902},
   number={7440},
   journal={Nature},
   publisher={Springer Science and Business Media LLC},
   author={Kirchmair, Gerhard and Vlastakis, Brian and Leghtas, Zaki and Nigg, Simon E. and Paik, Hanhee and Ginossar, Eran and Mirrahimi, Mazyar and Frunzio, Luigi and Girvin, S. M. and Schoelkopf, R. J.},
   year={2013},
   month=mar, pages={205–209} }

@article{zhu_circuit_2013,
	title = {Circuit {QED} with fluxonium qubits: Theory of the dispersive regime},
	volume = {87},
	url = {https://link.aps.org/doi/10.1103/PhysRevB.87.024510},
	doi = {10.1103/PhysRevB.87.024510},
	shorttitle = {Circuit {QED} with fluxonium qubits},
	pages = {024510},
	number = {2},
	journaltitle = {Physical Review B},
	shortjournal = {Phys. Rev. B},
	publisher = {American Physical Society},
	author = {Zhu, Guanyu and Ferguson, David G. and Manucharyan, Vladimir E. and Koch, Jens},
	urldate = {2025-07-18},
	date = {2013-01-14},
}

@book{Haroche,
      author = "Haroche, Serge and Raimond, Jean Michel",
      title = "{Exploring the Quantum}",
      publisher = "Oxford Univ. Press",
      address = "Oxford",
      year = "2006"
}

@ARTICLE{Thomas_cryo,
  author={Thomas, Jeremy N. and Hoffmann, Johannes and Flowers-Jacobs, Nathan E. and Fox, Anna E. and Jungwirth, Nicholas R. and Johnson-Wilke, Raegan L. and Dresselhaus, Paul D. and Benz, Samuel P.},
  journal={IEEE Transactions on Microwave Theory and Techniques}, 
  title={Cryogenic On-Chip In Situ S-Parameter Calibration Using Superconducting Coplanar Waveguides}, 
  year={2025},
  volume={73},
  number={11},
  pages={8942-8955},
  doi={10.1109/TMTT.2025.3585803}}

@article{Kimble_VRS,
  title = {Observation of normal-mode splitting for an atom in an optical cavity},
  author = {Thompson, R. J. and Rempe, G. and Kimble, H. J.},
  journal = {Phys. Rev. Lett.},
  volume = {68},
  issue = {8},
  pages = {1132--1135},
  numpages = {0},
  year = {1992},
  month = {Feb},
  publisher = {American Physical Society},
  doi = {10.1103/PhysRevLett.68.1132},
  url = {https://link.aps.org/doi/10.1103/PhysRevLett.68.1132}
}

@article{Kimble_mirrors,
  title = {Characterization of high-finesse mirrors: Loss, phase shifts, and mode structure in an optical cavity},
  author = {Hood, Christina J. and Kimble, H. J. and Ye, Jun},
  journal = {Phys. Rev. A},
  volume = {64},
  issue = {3},
  pages = {033804},
  numpages = {7},
  year = {2001},
  month = {Aug},
  publisher = {American Physical Society},
  doi = {10.1103/PhysRevA.64.033804},
  url = {https://link.aps.org/doi/10.1103/PhysRevA.64.033804}
}

@article{Miller_2005,
	author = {Miller, R and Northup, T E and Birnbaum, K M and Boca, A and Boozer, A D and Kimble, H J},
	doi = {10.1088/0953-4075/38/9/007},
	journal = {Journal of Physics B: Atomic, Molecular and Optical Physics},
	month = {apr},
	number = {9},
	pages = {S551},
	title = {Trapped atoms in cavity QED: coupling quantized light and matter},
	url = {https://doi.org/10.1088/0953-4075/38/9/007},
	volume = {38},
	year = {2005}}

@article{Lee_PhotonCal,
  title = {Ion-Based Quantum Sensor for Optical Cavity Photon Numbers},
  author = {Lee, Moonjoo and Friebe, Konstantin and Fioretto, Dario A. and Sch\"uppert, Klemens and Ong, Florian R. and Plankensteiner, David and Torggler, Valentin and Ritsch, Helmut and Blatt, Rainer and Northup, Tracy E.},
  journal = {Phys. Rev. Lett.},
  volume = {122},
  issue = {15},
  pages = {153603},
  numpages = {6},
  year = {2019},
  month = {Apr},
  publisher = {American Physical Society},
  doi = {10.1103/PhysRevLett.122.153603},
  url = {https://link.aps.org/doi/10.1103/PhysRevLett.122.153603}
}

@article{Youssefi_2023,
   title={A squeezed mechanical oscillator with millisecond quantum decoherence},
   volume={19},
   ISSN={1745-2481},
   url={http://dx.doi.org/10.1038/s41567-023-02135-y},
   DOI={10.1038/s41567-023-02135-y},
   number={11},
   journal={Nature Physics},
   publisher={Springer Science and Business Media LLC},
   author={Youssefi, Amir and Kono, Shingo and Chegnizadeh, Mahdi and Kippenberg, Tobias J.},
   year={2023},
   month=aug, pages={1697–1702} }

@article{Satzinger_2018,
   title={Quantum control of surface acoustic-wave phonons},
   volume={563},
   ISSN={1476-4687},
   url={https://doi.org/10.1038/s41586-018-0719-5},
   doi={10.1038/s41586-018-0719-5},
   number={7733},
   journal={Nature},
   publisher={Springer Science and Business Media LLC},
   author={Satzinger, K. J. and Zhong, Y. P. and Chang, H.-S. and Peairs, G. A. and Bienfait, A. and Chou, Ming-Han and Cleland, A. Y. and Conner, C. R. and Dumur, {\'E}. and Grebel, J. and Gutierrez, I. and November, B. H. and Povey, R. G. and Whiteley, S. J. and Awschalom, D. D. and Schuster, D. I. and Cleland, A. N.},
   year={2018},
   month=nov, pages={661–665} }

@article{quantumPhononicsReview, 
year = {2025}, 
title = {{Quantum Phononics: From Principles to Engineering}}, 
author = {Lei, Changyong and Wang, Zi and Yang, Chenwen and Liu, Yizhou and Ren, Jie}, 
journal = {The Journal of Physical Chemistry Letters}, 
issn = {1948-7185}, 
doi = {10.1021/acs.jpclett.5c00951}, 
pmid = {40689774}, 
pages = {7630--7641}, 
number = {30}, 
volume = {16}
}

@article{Weis_2010,
   title={Optomechanically Induced Transparency},
   volume={330},
   ISSN={1095-9203},
   url={http://dx.doi.org/10.1126/science.1195596},
   DOI={10.1126/science.1195596},
   number={6010},
   journal={Science},
   publisher={American Association for the Advancement of Science (AAAS)},
   author={Weis, Stefan and Rivière, Rémi and Deléglise, Samuel and Gavartin, Emanuel and Arcizet, Olivier and Schliesser, Albert and Kippenberg, Tobias J.},
   year={2010},
   month=dec, pages={1520–1523} }

@article{optomechanicsRevModPhys, 
year = {2014}, 
title = {{Cavity optomechanics}}, 
author = {Aspelmeyer, Markus and Kippenberg, Tobias J. and Marquardt, Florian}, 
journal = {Reviews of Modern Physics}, 
issn = {0034-6861}, 
doi = {10.1103/revmodphys.86.1391}, 
eprint = {1303.0733}, 
pages = {1391--1452}, 
number = {4}, 
volume = {86}
}

% \newpage
% \whitetext{blank page}
% \newpage

\clearpage
%\section*{End Matter}\label{sec:endmatter}

\begin{center}
    \large \textbf{End Matter}
    \vskip 0.2in
\end{center}

% \yxw{Let's put all the long equations in the supplement, only keep the method + (setup + final result for qubit and transmon)}

% \subsection{Fourth-order Schrieffer--Wolff perturbation theory}
% \label{appsec:SW4.gen}
\noindent
\textit{Fourth-order Schrieffer-Wolff perturbation theory}.---We formulate nondegenerate perturbation theory as a series of Schrieffer-Wolff transformations
	which successively diagonalize the Hamiltonian \(H=H_0+V\) to higher orders in the perturbation \(V\). We apply our formulation to compute self- and cross-Kerr for modes coupled via a qubit---first a two-level system and then a transmon---for all qubit levels, as operators, in closed form. Other works \cite{zhang_drive-induced_2022,Elliott_2018,zhu_circuit_2013} compute fourth-order corrections for modes coupled via a qubit but not for all qubit levels, in closed form.
    We work in a basis which diagonalizes \(H_{0} = \sum _{j} E_j {\ketbra{j}{j}}\) with the \(E_j\) the unperturbed eigenenergies and the \(\ket j\) the unperturbed eigenkets. We assume \(V\) is off-diagonal in this basis. The expansion is
\begin{align}
	%\prod_{n=1}^\infty \prod_{m=1}^\infty
    e^{i G_4} e^{i G_3} e^{i G_ 2} e^{i G_1} H e^{-i G_1} e^{-i G_2} e^{-i G_3} e^{-i G_4}
    \nonumber\\
	= \sum_{n=0}^4 H_n + O(V^5)
	\label{eq:SWseries}
\end{align}
where the \(G_n\) are \(n\)th-order Hermitian generators of the transformations and the \(H_n\) are diagonal corrections to the unperturbed Hamiltonian \(H_0\).

We obtain expressions for \(H_n\) up to \(n=4\). For each \(n\), we expand the left-hand-side of \eq{SWseries} to \(n\)th order using
\(
    e^{A} B e^{-A} = B + [A,B] + [A,[A,B]]/2! + \cdots
\)
and enforce the condition that \(H_n\) is diagonal by choosing \(G_n\) such that the \(n\)th-order term \(i[G_n,H_0]\), which is off-diagonal, exactly cancels the off diagonal part of the remaining \(n\)th-order terms. 

We first recall the textbook example of first-order perturbative correction.
Equating first-order parts of either side of \eq{SWseries} gives
\(
	i[G_1, H_0]+V = H_1.
	%\label{eq:SWseriesO1}
\)
\(H_1\) must be diagonal so we choose \(G_1\) such that the off-diagonal part of \(i[G_1, H_0]+V\) is zero. We thus have
\begin{align}
    \bra j G_1 \ket k &= -\frac {i \bra j V \ket k}{E_j - E_k}\ \text{for}\ j \neq k. 
    \label{eq:G1}
\end{align}
Note that in order for the generators to be well-defined, the spectrum of \(H_0\) must be nondegenerate.
The diagonal elements of \(G_1\) are undetermined and we are free to choose
\(
    \bra j G_1 \ket j = 0.
\)
Since by assumption \(V\) is off-diagonal we have
\(
    H_1 = 0.
\)
\(G_1\) does not appear in the final expression for \(H_1\), but does in those for \(H_n\) with \(n\geq2\).
The fact that \(i[G_1, H_0]+V=H_1=0\) can be used to reduce the number of terms in the expansion of the left-hand-side of \eq{SWseries}.

To make the expressions for higher-order corrections more compact, we define notation for the diagonal part of an operator in the \(H_0\) eigenbasis:
\(
    \diag A = \sum_j \ket j \bra j A \ket j \bra j.
\)

Repeating the process at second order yields
\begin{align}
    H_2 &= i[G_2,H_0]+\frac {i [G_1,V]} 2= \frac{i \diag[G_1,V]} 2,
    \\
    \bra j G_2 \ket k &= \frac{\bra j [G_1,V] \ket k}{2 (E_j - E_k)}\ \text{for}\ j\neq k.
\end{align}
As with \(G_1\), the diagonal elements of \(G_2\) are undetermined and we choose
\(
    \diag G_2=0.
\)

Developing the expansion further we find the third-order correction
\begin{align}
    H_3 &= -\frac{\diag [G_1, [G_1, V]]} 3,
\end{align}
and the fourth-order correction
\begin{align}
    H_4
    &= -\diag \bigg( \frac{i [G_1,[G_1,[G_1,V]]]}{8} + \frac{ [ G_2, [G_1, V]]} 4 \bigg).
    \label{eq:H4}
\end{align}

In following sections we apply \eq{H4} to compute Kerr coefficients for two modes each coupled to a qubit. First for a two-level system, which is the simplest scenario and then for a transmon which is more realistic.

\textit{Cross- and self-Kerr terms for two modes coupled via a two-level system}.---%
Let \(\sigma^\dagger = \ket 1 \bra 0\) be the spin-raising operator for a two-level system with ground state \(\ket0\) and excited state \(\ket 1\) and \(\sigma_z = \ket 1 \bra 1 - \ket 0 \bra 0\). We model two modes coupled via a two-level system with the Hamiltonian \(H = H_0 + V\), where
\begin{align}
    H_0 = \omega_\aLabel a^\dagger a + \omega_\bLabel b^\dagger b + \omega_q \frac {\sigma_z} 2
\end{align}
and
\begin{align}
    V = g_\aLabel (a \sigma^\dagger + a^\dagger \sigma)+ g_\bLabel(b \sigma^\dagger+b^\dagger \sigma)
\end{align}
is the sum of Jaynes--Cummings-type couplings between each mode and the two-level system in the rotating-wave approximation. Here, \(a^\dagger\) and \(b^\dagger\) are bosonic creation operators for photons of frequency \(\omega_\aLabel\) and \(\omega_\bLabel\), respectively, and \(\sigma = \ket 0 \bra 1\) is the spin lowering operator for the two-level system. Perturbation theory is valid if the detunings are in the dispersive regime:
\(
    |\Delta_\aLabel|,\,|\Delta_\bLabel|\gg g_\aLabel %\sqrt{\langle a^\dagger a\rangle+1}
    ,\,g_\bLabel %\sqrt{\langle b^\dagger b \rangle+1}
    ,
\)
where \(\Delta_{\aLabel(\bLabel)} = \omega_q -\omega_{\aLabel(\bLabel)}\) is the detuning of the two-level system from mode \(\aLabel\ (\bLabel)\).

We apply the general formalism developed above to this Hamiltonian. The first-order generator is
\begin{align}
    G_1 = -i g_\aLabel \bigg(\frac{a\sigma^\dagger}{\Delta_\aLabel} - \frac{a^\dagger \sigma}{\Delta_\aLabel}\bigg)
    -i g_\bLabel \bigg(\frac{b\sigma^\dagger}{\Delta_\bLabel} - \frac{b^\dagger \sigma}{\Delta_\bLabel}\bigg).
\end{align}
To obtain \(H_2\), we evaluate \(i[G_1,V]/2\).
Collecting the diagonal (in the $H_0$ eigenbasis) terms and dropping a constant, we obtain the second-order correction
\begin{align}
    H_2 = \bigg[
		\frac{g_\aLabel^2}{\Delta_\aLabel}\bigg(a^\dagger a + \frac 1 2\bigg)
		+\frac{g_\bLabel^2}{\Delta_\bLabel}\bigg(b^\dagger b + \frac 1 2\bigg)
	\bigg] \sigma_z,
    \label{eq:TLSH2}
\end{align}
which is just two copies of the dispersive Hamiltonian for a bosonic mode coupled to a two-level system. Mode--mode interaction does not appear at second order.

The third-order correction vanishes, i.e.~\(H_3=0\).
This follows from the fact that the Hamiltonian is invariant under the transformation
	\( (g_\aLabel,g_\bLabel,a,b)\to-(g_\aLabel,g_\bLabel,a,b)\).
This argument implies \(H_n = 0\) for all odd \(n\).
The second-order generator is
\begin{align}
    G_2 = - \frac{i g_\aLabel g_\bLabel}{2 (\Delta_\aLabel - \Delta_\bLabel)}\bigg( \frac 1 {\Delta_\aLabel} + \frac 1 {\Delta_\bLabel} \bigg)(a^\dagger b - a b^\dagger)\sigma_z.
\end{align}
The full fourth-order correction is
\begin{align}
\label{eq:TLSunsimplified.sum}
    H_4 &= \omega_\aLabel^{(4)}  a^\dagger a + \omega^{(4)}_\bLabel b^\dagger b +\omega_q^{(4)} \sigma^\dagger \sigma
    \nonumber\\&\qquad
    + \chi_\aLabel^{(4)} a^\dagger a \sigma^\dagger \sigma + \chi_\bLabel^{(4)} b^\dagger b \sigma^\dagger \sigma
        \nonumber\\&\qquad
        + \frac 1 2 O_{\aLabel\aLabel} a^\dagger a^\dagger a a + \frac 1 2 O_{\bLabel\bLabel} b^\dagger b^\dagger bb
        \nonumber\\&\qquad
         +  O_{\aLabel\bLabel} a^\dagger a b^\dagger b,
\end{align}
where
\begin{align}
    O_{\aLabel\aLabel} &= -\frac{2 g_\aLabel^4}{\Delta_\aLabel^3} \sigma_z,
    \label{eq:TLSsk}
    \\
    O_{\aLabel\bLabel} &= -\frac{2 g_\aLabel^2 g_\bLabel^2}{\Delta_\aLabel \Delta_\bLabel} \bigg(\frac 1 \Delta_\aLabel + \frac 1 \Delta_\bLabel\bigg) \sigma_z.
    \label{eq:TLSck}
\end{align}
Projecting into the ground state of the two-level system gives
\begin{align}
\chi _{\aLabel\aLabel} = \bra0 O_{\aLabel\aLabel}\ket0 &= \frac{2 g_\aLabel^4 }{\Delta_\aLabel^3}
\label{eq:groundTLSsk}
,
\\
\chi _{\aLabel\bLabel} = \bra0 O_{\aLabel\bLabel} \ket0 &= \frac{2 g_\aLabel^2 g_\bLabel^2}{\Delta_\aLabel \Delta_\bLabel}\bigg(\frac 1 \Delta_\aLabel + \frac 1 \Delta_\bLabel \bigg)
\label{eq:groundTLSck}
.
\end{align}
Eqs.~(\ref{eq:groundTLSsk},\,\ref{eq:groundTLSck}) are a first approximation for the self- and cross-Kerr induced on two modes coupled by a qubit. In the next section, we increase the complexity and accuracy of our model by replacing the two-level system with a transmon.

\textit{Cross- and self-Kerr terms for two modes coupled via a transmon.}---The Hamiltonian for two modes coupled via a transmon in the rotating-wave approximation is \(H=H_0+V\) with
\begin{align}
	H_{0} &= \omega_\aLabel a^\dagger a 
		+ \omega_\bLabel b^\dagger b
		+ \omega_q c^\dagger c - \frac \alpha 2 c^\dagger c^\dagger c c
\end{align}
and
\begin{align}
    V&= g_\aLabel (c^\dagger a + c a^\dagger) + g_\bLabel (c^\dagger b + c b^\dagger).
\end{align}
Here, \(a^\dagger\) and \(b^\dagger\) are bosonic
	creation operators for photons of frequency \(\omega_\aLabel\) and \(\omega_\bLabel\), respectively, 
\(c^\dagger\) is the bosonic creation operator for the transmon, and 
\(\alpha\) is the transmon's anharmonicity.
Each mode is coupled to the transmon at rate \(g_\aLabel\) (\(g_\bLabel\)), but the modes are not directly coupled to each other.
The transmon mediates an interaction between the modes which does not appear until the fourth-order of perturbation theory. As for the two modes and two-level system, perturbation theory is valid in the dispersive regime.

The first-order generator now reads [cf. Eq.~\eqref{eq:G1}]
\begin{align}
	G_1 &= - i g_\aLabel a c^\dagger \frac 1 {\Delta_\aLabel-\alpha c^\dagger c}
    - i g_\bLabel b c^\dagger \frac 1 {\Delta_\bLabel-\alpha c^\dagger c} + \text{H.c.}.
\end{align}
This operator includes rational functions of the transmon number operator \(c^\dagger c\)
because the denominator in \eq{G1} depends on the occupation number of the transmon.

We introduce notation \(x+\{\aLabel\leftrightarrow \bLabel\}\) which means the sum of $x$ and its permutation where all quantities associated with mode \(\aLabel\) are exchanged with all quantities associated with mode \(\bLabel\).
The second-order correction reads
\begin{align}
	H_2 &= -g_\aLabel^2 \Bigg[ 
            a^\dagger a \Bigg(\frac 1 {\Delta_\aLabel-\alpha c^\dagger c}
            + c^\dagger \frac 1 {\Delta_\aLabel-\alpha (c^\dagger c+1)} c\Bigg)
            \nonumber\\&\qquad
            -(a^\dagger a + 1) c^\dagger \frac 1 {\Delta_\aLabel-\alpha c^\dagger c} c
        \Bigg]
        +\{\aLabel \leftrightarrow \bLabel\}
        .
\end{align}

The third-order correction vanishes: \(H_3=0\).
Again, this follows from the fact that the Hamiltonian is invariant under the transformation \((g_\aLabel,g_\bLabel,a,b)\to-(g_\aLabel,g_\bLabel,a,b)\).
This argument implies \(H_n = 0\) for all odd \(n\).
For the sake of compactness, we define the operators
\begin{align}
    f_{i,\mu} &= \frac 1 {\Delta_i-\alpha(c^\dagger c + \mu)},
    \label{eq:f_def}
    \\
    h_{ij,\mu} &= \frac 1 {\Delta_i + \Delta_j - \alpha(2 c^\dagger c + 1+\mu)}.
    \label{eq:h_def}
\end{align}
The second-order generator reads
\begin{align}
	G_2 &= \frac 1 2 [
        - i g_\aLabel^2 a a c^\dagger c^\dagger ( 
            f_{A,1}-f_{A,0}
        )
        h_{\aLabel\aLabel,0}
        \nonumber\\
        &\qquad- i g_\aLabel g_\bLabel a b c^\dagger c^\dagger
        ( 
            f_{B,1}-f_{B,0}
        )
        h_{\aLabel\bLabel,0}
        \nonumber\\
        &\qquad
        + i g_\aLabel g_\bLabel a b^\dagger (
            f_{B,0} + c^\dagger [
                f_{B,1}-f_{B,0}
            ] c 
        ) \frac 1 {\Delta_\aLabel - \Delta_\bLabel} 
        \nonumber\\
        &\qquad
        + \{\aLabel\leftrightarrow \bLabel\}
    ] + \text{H.c.}
\end{align}
The fourth-order correction has the form
\begin{align}
    H_4 &= \omega_\aLabel^{(4)}  a^\dagger a + \omega^{(4)}_\bLabel b^\dagger b +O_q
        + O_\aLabel a^\dagger a + O_\bLabel b^\dagger b
        \nonumber\\&\qquad
        + \frac 1 2 O_{\aLabel\aLabel} a^\dagger a^\dagger a a + \frac 1 2 O_{\bLabel\bLabel} b^\dagger b^\dagger bb
        \nonumber\\&\qquad
        + O_{\aLabel\bLabel} a^\dagger a b^\dagger b.
    \label{eq:transmonFourthOrder}
\end{align}
\(O_{\aLabel\aLabel}\) is the self-Kerr for mode \(\aLabel\) and \(O_{\aLabel\bLabel}\) is the cross-Kerr between mode \(\aLabel\) and mode \(\bLabel\). 
Full expressions for \(O_{\aLabel\aLabel}\) and \(O_{\aLabel\bLabel}\)---self- and cross-Kerr for two modes coupled via a transmon for all levels of the transmon, in closed form---appear in the Supplemental Material and are, to our knowledge, a new result. 
The Kerr coefficients for the transmon in the ground state are already known \cite{zhang_drive-induced_2022,Elliott_2018}.
Projecting into the transmon ground state gives
\begin{align}
    \chi_{\aLabel\aLabel} &= \bra0O_{\aLabel\aLabel}\ket0
    =-\frac{2\alpha g_\aLabel^4}{\Delta_\aLabel^3 (2\Delta_\aLabel-\alpha)},
    \label{eq:groundTransmonSK}
    \\
    \chi_{\aLabel\bLabel}&= \bra0O_{\aLabel\bLabel}\ket0
    \nonumber\\
    &=-\frac{2 \alpha g_\aLabel^2 g_\bLabel^2}{\Delta_\aLabel \Delta_\bLabel (\Delta_\aLabel+\Delta_\bLabel-\alpha)} \bigg( \frac 1 {\Delta_\aLabel} + \frac 1 {\Delta_\bLabel} \bigg),
    \label{eq:groundTransmonCK}
\end{align}
[c.f., Eqs.~(\ref{eq:SelfKerrMain},\,\ref{eq:CrossKerrMain}) in the main text]. In the limit \(\alpha\to\infty\), we recover Eqs.~(\ref{eq:groundTLSsk},\,\ref{eq:groundTLSck}), the result for a two-level system. In the limit \(\alpha\to0\) we recover linearity (i.e., \(\chi_{\aLabel\aLabel}\ \text{and}\ \chi_{\aLabel\bLabel}\to0\)).

\newpage

\clearpage
\onecolumngrid

% Re-enable TOC writing
\makeatletter
\let\addcontentsline\latex@addcontentsline
\makeatother

% \section*{}
\begin{center}
    \large \textbf{Supplemental Material: In-situ Characterization of Light-Matter Coupling in Multimode Circuit-QED Systems}
    \vskip 0.1in

\end{center}

\vskip 0.3in

% for supplementary material naming of figures and tables
\setcounter{figure}{0}
\setcounter{equation}{0}
\setcounter{section}{0}
\renewcommand{\figurename}{Figure}
\renewcommand{\thefigure}{S\arabic{figure}}
\renewcommand{\thetable}{S\arabic{table}}
\renewcommand{\theequation}{S\arabic{equation}}

\renewcommand{\thesection}{S\arabic{section}}
\tableofcontents

% %%%%trying to have table to contents only for appendices. AK 2/18/26
% \addcontentsline{toc}{section}{Appendix} % Add the appendix text to the document TOC
% % \part{Appendix} % Start the appendix part
% \parttoc % Insert the appendix TOC

\section{CPW Lattice Device}\label{app:Device}

\begin{figure*}[t]
\centering
    \includegraphics[width=\textwidth]{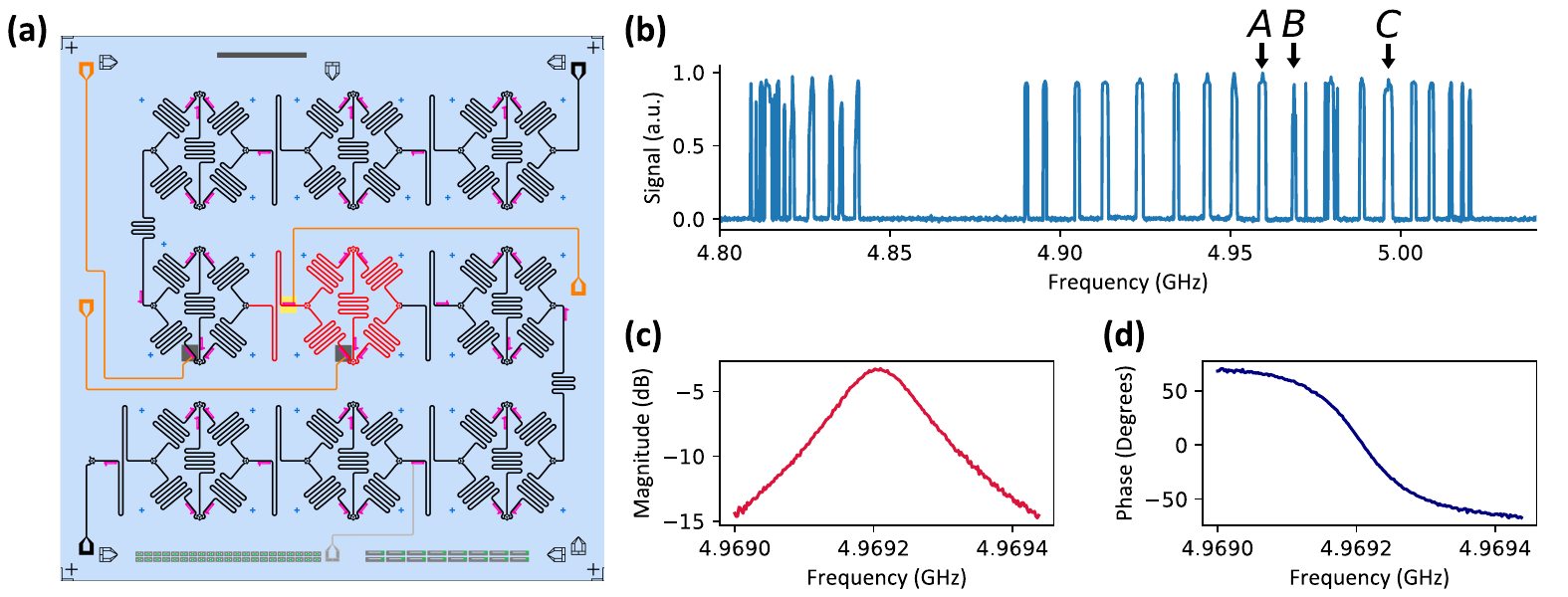}
	% \vspace{-0.6cm}
    \vspace{-0.2cm}
    \caption{\label{fig:mode_spectroscopy} 
    \textit{Device layout and photonic modes.} (a) CAD of the CPW lattice device used in this work with colors added to highlight details. The lattice resonators are shown in black, with the exception of the resonators in the central unit cell which are shown in red. Pockets created in the side of each CPW where qubits can be added are shown in pink. The three embedded flux-tunable transmon qubits are highlighted with squares. The flux-tunable transmon used to collect the data in this work is highlighted with a yellow square and the other two transmons are highlighted with gray squares. Their corresponding flux-bias lines are shown in orange. Input/output ports in the bottom-left and top-right corners of the device allow microwave transmission measurements. (b) High-power two-tone spectroscopy measurement of the CPW lattice spectrum (a detailed description of this measurement technique can be found in Ref.~\cite{OBrien2025}). The three modes used to test the coupling verification protocol are indicated with gray arrows ($\omega_A/2 \pi = 4.960$~GHz, $\omega_B/2 \pi = 4.969$~GHz, $\omega_C/2 \pi = 4.997$~GHz). (c)-(d) Transmission spectroscopy of mode $B$ showing the magnitude and phase of the transmitted microwave tone respectively as a function of signal frequency. A correction is applied to the measured phase response of the mode to account for linear phase delay from the transmission line.} 
  
\end{figure*}

The data presented in this work were taken on the quasi-1D coplanar-waveguide (CPW) resonator lattice device with embedded transmon qubits shown in Fig.~\ref{fig:mode_spectroscopy}(a).
The CPW resonators are defined lithographically on a $200$~nm film of tantalum coated onto a $25.4 \times 25.4\ \mathrm{mm}^2$ wafer of sapphire substrate.
The lattice consists of nine unit cells containing six resonators each; the first unit cell is coupled to an input port while the last is coupled to an output port.
All resonators in the device are end-to-end coupled to nearest neighbors through 3-way capacitors.
Three frequency-tunable transmon qubits are capacitively coupled to individual resonators within the lattice.
The capacitor paddles of the transmons are defined lithographically in the tantalum film alongside the CPW resonators while the Josephson junctions, composed of two layers of aluminum separated by an insulating layer of aluminum oxide, are deposited onto the chip through double-angle evaporation.
The data in this work utilize only one of the three transmons, highlighted with a yellow square in Fig.~\ref{fig:mode_spectroscopy}(a).
The other two transmons are kept detuned $\sim 5$~GHz from the relevant normal modes to ensure that their contributions to the observed Kerr shifts are negligible.
A more in-depth description of the device, including details of its fabrication and circuit parameters, can be found in Ref.~\cite{OBrien2025}.

Though the resonators in the device are all designed to have the same frequency, the nearest-neighbor capacitive coupling lifts this degeneracy.
CPW resonator arrays in the absence of qubits can approximately be described by a tight-binding Hamiltonian,
\begin{equation}\label{eq:TB_Model}
H =  \,  \omega_0 \sum_{n} a_n^{\dagger} a_n - \, t \sum_{\langle n, n' \rangle} a_n^{\dagger}a_{n'},
\end{equation}
where $\omega_0$ is the resonator frequency, primarily set by the length of the CPW, and $t$ is the effective hopping rate between coupled resonators, primarily set by the coupling capacitance.
The bosonic raising/lowering operators for each resonator, $a^{\dagger}$/$a$, are indexed by the integer $n$, and $\langle n, n' \rangle$ indicates neighboring resonators that share a coupler.
Consequently, the frequencies of the normal modes are highly dependent on the coupling geometry.
The mode spectrum of the quasi-1D lattice device, measured with high-power two-tone spectroscopy using a sensor mode at $4.960$~GHz, is shown in Fig.~\ref{fig:mode_spectroscopy}(b).

\section{Two-Tone Spectroscopy}\label{app:TwoTone}

\begin{figure*}[t]
\centering
    \includegraphics[width=1\textwidth]{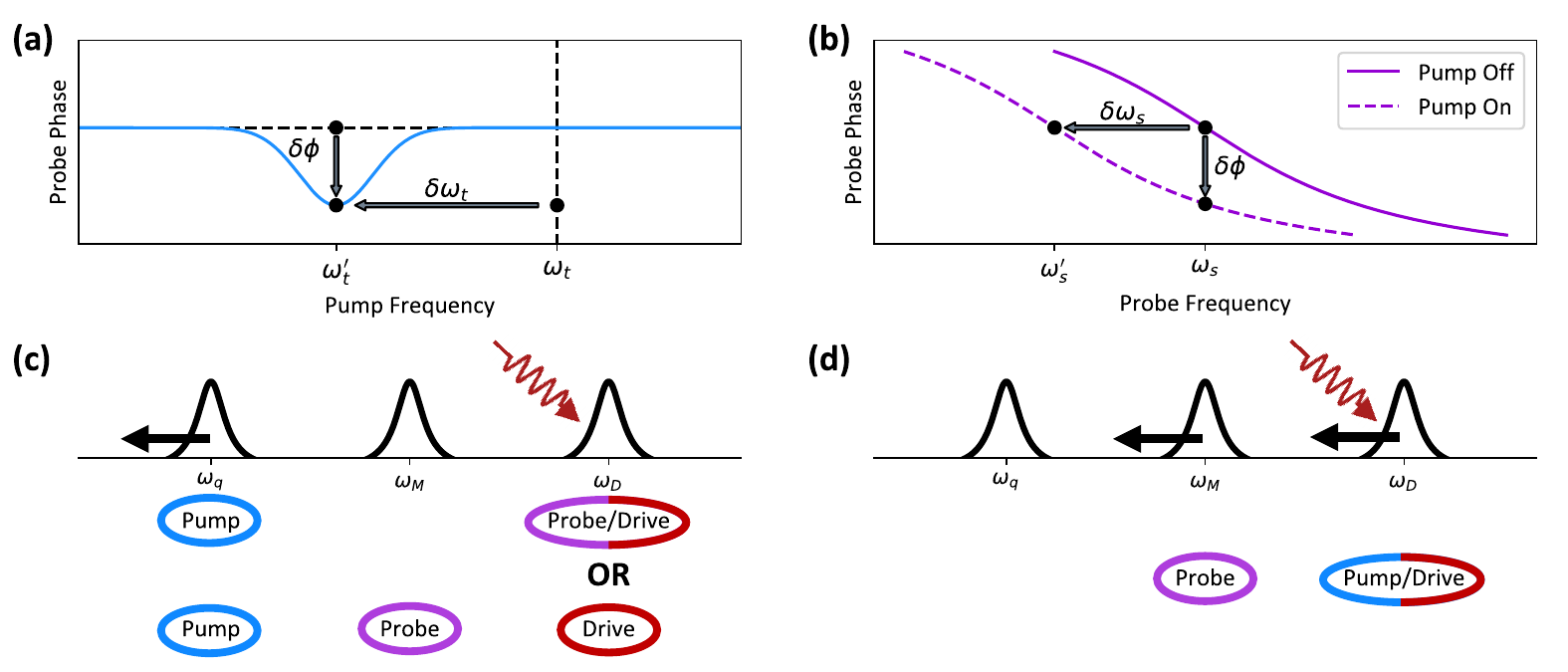}
	% \vspace{-0.6cm}
    \vspace{-0.2cm}
    \caption{\label{fig:two_tone} 
    \textit{Generalized two-tone spectroscopy.} (a) Schematic depicting general two-tone spectroscopy measurements. The phase of a probe tone transmitted through the sensor mode is measured while the frequency of a pump tone is swept over a target feature, either a qubit or a second normal mode. The strongest phase response $\delta \phi$ occurs when the pump frequency is resonant with the effective frequency of the target $\omega_t'$, indicating that the target frequency is shifted by $\delta \omega_t$ from its original frequency $\omega_t$. (b) Schematic depicting the phase response function of the sensor mode to a probe tone for two cases: (i) where no pump tone is applied to the target feature (solid line), and (ii) where a pump tone resonant with the target feature is applied (dashed line), shifting the center frequency of the probed mode by $\delta \omega_s$ from its original value $\omega_s$. As the expected response of the sensor is known prior to the measurement, the phase difference $\delta \phi$ observed in two-tone spectroscopy is used to determine $\delta \omega_s$. (c) Schematic of the AC-Stark-shift measurement highlighting the roles of the microwave tones involved. Two variants of the two-tone spectroscopy measurement used to track the effective qubit frequency are shown: (i) the method used to collect the data presented in this work, and (ii) a modification for systems where the transmitted signal through the drive mode is negligible. In the first case, mode $D$ functions as the sensor mode, and the probe tone through the mode doubles as the drive producing the AC-Stark shift on the qubit frequency. In the second, mode $M$ functions as the sensor mode; a probe tone is sent through mode $M$ and a dedicated drive tone is applied to mode $D$. In both cases, the qubit is the target feature and is excited by a pump tone. (d) Schematic of the Kerr-shift measurement highlighting the roles of the microwave tones involved. A probe tone is sent on resonance with mode $M$, which functions as the sensor mode. Mode $D$ functions as the target feature and the frequency of a pump tone is swept over the mode. This pump tone doubles as the drive producing the self-Kerr shift on mode $D$ and the cross-Kerr shift on mode $M$.} 
  
\end{figure*}

This section contains a more detailed description of two-tone spectroscopy, the scanning pump-probe technique used throughout the measurement protocol for determining qubit-mode coupling.
Though two-tone spectroscopy is a common technique in the field of circuit QED to perform qubit measurements \cite{Blais:revmodphys}, the methods used in the Kerr measurement to probe other nonlinear modes are less common and warrant further discussion \cite{Peugeot_2024, Bosman_2017}.
To ensure that the following description of two-tone spectroscopy is completely general and applies to both the AC-Stark-shift and Kerr-shift measurements, we introduce two new terms: the sensor mode and the target feature. (The connection between the more general target/sensor terminology and the drive/monitor terminology used in the main text is shown in Fig.~\ref{fig:two_tone} and Section~\ref{app:Protocol_Details}.)

During the measurement, a probe tone is sent in on resonance with a sensor mode while the frequency of a pump tone is swept in the vicinity of a target feature, either a qubit or a different photonic mode.
When the pump tone populates the target feature with photons, the sensor mode is shifted off resonance from the probe, modifying the transmission of the probe through the device.
Through measurements of the transmitted probe signal versus pump frequency, the frequency of the target feature can be located.

To explain this process, we first define more explicitly the relationship between the sensor mode and the target feature.
The effective frequency of the sensor mode is given by
\begin{equation}\label{eq:effective_mode_freq}
\omega_{s}' = \omega_{s} + \chi n_{t} = \omega_s + \delta\omega_s,
\end{equation}
where $\omega_s$ is the unperturbed frequency of the sensor and $n_t$ is the average photon number in the target feature.
Depending on the nature of the target feature, the parameter $\chi$ is either the dispersive shift from a transmon or the cross-Kerr coefficient from a second normal mode.
We write the change in the sensor mode frequency as $\delta \omega_s \equiv \omega_s'-\omega_s = \chi n_t$.
For simplicity, the self-Kerr shift on the sensor frequency from probe photons in the sensor mode is omitted from Eq.~(\ref{eq:effective_mode_freq}).

Consequently, the transmission of the probe tone through the sensor is dependent on whether the pump has populated the target feature with photons.
The sensor response to the probe is a function of $\omega_{\text{pr}}$ and $\omega_s')$ that we denote as $\mathcal{R}(\omega_{\text{pr}}, \omega_s')$.
In particular, in this manuscript, we use the transmission response of the sensor mode.
As the response depends only on the detuning between the probe and effective sensor frequency, it can be rewritten as $R(\omega_s'-\omega_{\text{pr}})$.
%The sensor response to the probe, which depends only on the detuning between the probe frequency and the effective sensor frequency, is given by $\mathcal{R}(\omega_{\text{pr}}, \omega_s') = R(\omega_s'-\omega_{\text{pr}})$.
As the probe is resonant with the unperturbed sensor mode ($\omega_{\text{pr}} = \omega_s$), this simplifies even further to $R(\delta\omega_s)$, depending only on the frequency shift of the sensor.
When the pump is resonant with the target, $\omega_{\text{pump}} = \omega_t$, both the photon number in the target feature ($n_t$) and the sensor frequency shift ($\delta \omega_s$) are maximized, producing the largest change in the sensor response, $\vert\delta R \vert = \vert R(\delta \omega_s) - R(0) \vert$.
Thus, the frequency of the target is determined spectroscopically by finding the pump frequency which produces the largest change in the transmission of the probe.

By the same logic, two-tone spectroscopy can identify the effective frequency of the target feature if the target is experiencing an AC-Stark or Kerr shift. The details of how we induce AC-Stark and Kerr shifts will be described in Fig.~\ref{fig:two_tone} and Section~\ref{app:Protocol_Details}. For now we will consider the abstract scenario of a target feature with a frequency shift $\delta \omega_t$ such that the effective frequency of the target feature is given by
\begin{equation} \label{eq:eff_target_freq}
    \omega_t' = \omega_t+\delta \omega_t.
\end{equation}
% The effective target frequency is given by
% \begin{equation} \label{eq:eff_target_freq}
%     \omega_t' = \omega_t+\delta \omega_t,
% \end{equation}
% with the target feature experiencing a frequency shift of $\delta \omega_t$.
The highest photon population of the target feature (and thus the largest change in the sensor response) occurs when $\omega_{\text{pump}} = \omega_t'$, meaning the effective frequency of the target can also be tracked through this method.
The process of identifying the target frequency from the transmission of the probe is illustrated in Fig.~\ref{fig:two_tone}(a).

If the sensor response function is known, the frequency shift $\delta \omega_s$ on the sensor can be extracted from observations of $\vert\delta R \vert$.
The photonic mode response $R(\delta \omega_s)$ is a complex-valued function that specifies both the amplitude and phase of the output signal.
%It is simpler to work with the amplitude or phase response alone, both of which are real-valued expressions.
As the magnitude response is flat near resonance, we here consider only the phase of the transmission response, which is given by~\cite{steckquantum} 
\begin{equation}\label{eq:phase_response}
\phi(\omega_s'-\omega_{pr}) = \arctan{[2 Q(\omega_s'-\omega_{\text{pr}})/\omega_s']},
\end{equation}
where $Q$ is the quality factor of the sensor mode.
Considering again the case of a resonant probe, $\omega_{\text{pr}} = \omega_s $, the phase response simplifies to $\phi(\delta \omega_s) = \arctan{[2 Q \delta \omega_s/\omega_s']}$.
By measuring the probe phase during two-tone spectroscopy and inverting this expression, the frequency shift $\delta \omega_s$ on the sensor induced by the pump tone can be extracted.
A schematic displaying the phase response 
is shown in Fig.~\ref{fig:two_tone}(a) and (b).

%\ko{Maybe put the magnitude Lorentzian response in as well?}

\section{Experimental Measurement Protocol}\label{app:Protocol_Details}

Here we describe the full qubit-mode-coupling measurement protocol in detail.
In particular, we describe the measurements of the AC-Stark and Kerr shifts as extensions of general two-tone spectroscopy measurements and provide a more in-depth explanation of the models used to fit the data. 
Additionally, we discuss the measurements used to characterize the response function of the photonic modes.

To begin, the high-level description of the protocol as described in the main text is briefly reiterated.
Two modes, a monitor mode ($M$) and a drive mode ($D$), which are both coupled to the transmon qubit, are selected from the full device spectrum. 
In the first measurement, a microwave drive applied to $D$ with varying drive power $P$ populates the mode with photons, and the resultant AC-Stark shift per unit power on the frequency of the transmon, $\frac{\partial \omega_q'}{\partial P} = \beta \acd$, is observed.
In the second, a microwave drive with stronger drive power $P$ is again applied to mode $D$ and both the self-Kerr shift per unit power on the frequency of mode $D$, $\frac{\partial \omega_D'}{\partial P} = \beta \chi_{DD}$, and the cross-Kerr shift per unit power on the frequency of mode $M$, $\frac{\partial \omega_M'}{\partial P} = \beta \chi_{DM}$, are tracked.
The ratios $\beta \acd/ \beta \chi_{DD}$ and $\beta \acd/\beta \chi_{DM}$, which remove the dependence on the unknown power to photon number conversion factor $\beta$ for the drive mode, are used in conjunction with Eqs.~(\ref{eq:chi_main}-\ref{eq:CrossKerrMain}) to determine the qubit-mode couplings with the drive and monitor, $g_D$ and $g_M$.

The individual steps of the protocol are now described in detail.
Prior to the AC-Stark-shift and Kerr-shift measurements, transmission spectroscopy measurements are performed on modes $D$ and $M$ to identify $\omega_D$ and $\omega_M$, and to calibrate the phase response of the monitor mode.
% The magnitude response of a normal mode's transmission is described by a Lorentzian centered around the mode frequency.
% By determining the signal frequency that results in maximum transmission, the mode frequency is identified.
% The phase response function of mode $M$ is also characterized through a transmission measurement.
% The phase of the transmitted signal is recorded and, 
Following a linear phase correction to account for phase shifts introduced by the length of the input/output transmission lines, the phase of the transmitted signal is fit to Eq.~(\ref{eq:phase_response}).
This fitted function is later used in the Kerr-shift measurement to convert the phase of a probe tone resonant with mode $M$ to the magnitude of the cross-Kerr shift experienced by that mode. 
The measured magnitude and phase responses for a lattice mode at $4.9692$~GHz are shown in Fig.~\ref{fig:mode_spectroscopy}(c) and (d).

\begin{figure*}[t]
\centering
    \includegraphics[width=1\textwidth]{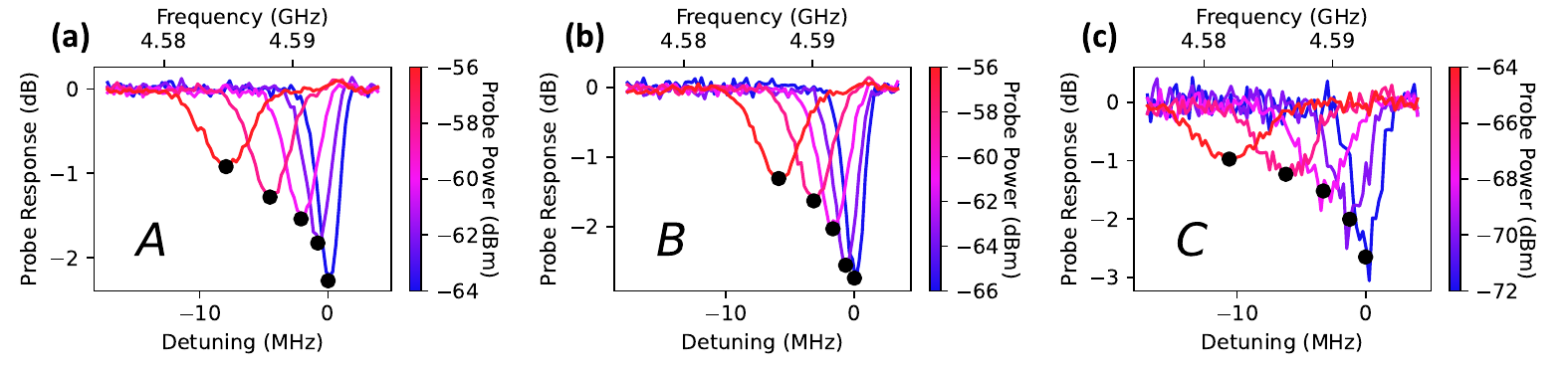}
	% \vspace{-0.6cm}
    \vspace{-0.2cm}
    \caption{\label{fig:stark_full} 
    \textit{AC-Stark shift.} (a) Two-tone spectroscopy of the transmon using mode $A$ as a sensor. The measurement is repeated for different drive powers ($P$) to observe the AC-Stark shift per unit power. Sample data from five drive powers are shown. The qubit frequency for each power, as determined by a Gaussian fit, is marked with a black dot. (b)-(c) Analogous measurements for a microwave drive applied to modes $B$ and $C$ respectively. For ease of comparison with Fig.~\ref{fig:raw_kerr_supp}, which shows Kerr measurements for all drive/monitor mode pairs, the drive mode (A,B or C) for these AC-Stark measurements is indicated inside each subpanel. The frequencies of the three drive modes, and their locations within the bands of the lattice are shown in Fig.~\ref{fig:mode_spectroscopy}(b).} 
  
\end{figure*}

We now describe the two-tone spectroscopy components of the protocol, starting with the characterization of the AC-Stark shift.
Using the terminology for general two-tone spectroscopy defined in Section~\ref{app:TwoTone}, mode $D$ acts as the sensor and the qubit acts as the target feature.
Accordingly, the magnitude of a probe tone transmitted through mode $D$ is measured while the frequency of a pump tone is swept over the qubit.
For the data collected in this work, the probe tone simultaneously functions as the drive that populates mode $D$ with photons to produce an AC-Stark shift and the power $P$ of the probe/drive tone is varied between measurements.
%the power of the probe tone is equivalent to the drive power $P_D$ and is varied between measurements.
%When the pump excites the qubit, the sensor frequency shifts and there is a decrease in the transmitted magnitude of the probe.
The pump frequency which produces the largest change in the transmission of the probe corresponds to the effective qubit frequency $\omega_q'$ and is identified for each value of $P$ by taking the center of a Gaussian fit to the measured transmission of the probe.
Sample data for the AC-Stark-shift measurement are shown in Fig.~\ref{fig:stark_full}.

The AC-Stark-shift measurement described above is effective for systems where the response of mode $D$ is easily observable in transmission spectroscopy, which is not guaranteed to be the case.
For example, in CPW lattice devices, there can exist highly-localized modes with weak coupling to the input or output ports which are not observable in transmission \cite{OBrien2025}. 
For cases where mode $D$ is not observable, the measurement can be easily modified to use mode $M$ as a sensor instead.
In this configuration, a weak probe is transmitted through mode $M$, a pump is used to excite the qubit, and a third microwave tone that functions as the drive is applied on resonance with mode $D$.
The power $P$ of this third tone is varied between measurements to produce the desired AC-Stark shifts, and the procedure to identify $\omega_q'$ using a sensor is carried out in the same manner as above.
Both of the possible microwave tone combinations for carrying out the AC-Stark-shift measurement are illustrated in Fig.~\ref{fig:two_tone}(c).

To carry out the Kerr-shift measurement with two-tone spectroscopy, mode $M$ is used as the sensor mode and mode $D$ acts as the target feature.
The phase of a probe tone transmitted through mode $M$ is observed while the frequency of a pump tone is swept over mode $D$.
Here, the pump tone doubles as the drive populating mode $D$ with photons.
The power $P$ of the pump/drive tone is varied between measurements.
When the pump populates mode $D$ with photons, it cross-Kerr shifts the frequency of mode $M$ off resonance from the probe, resulting in a change in the phase of the transmitted signal.

\begin{figure*}[t]
\centering
    \includegraphics[width=1\textwidth]{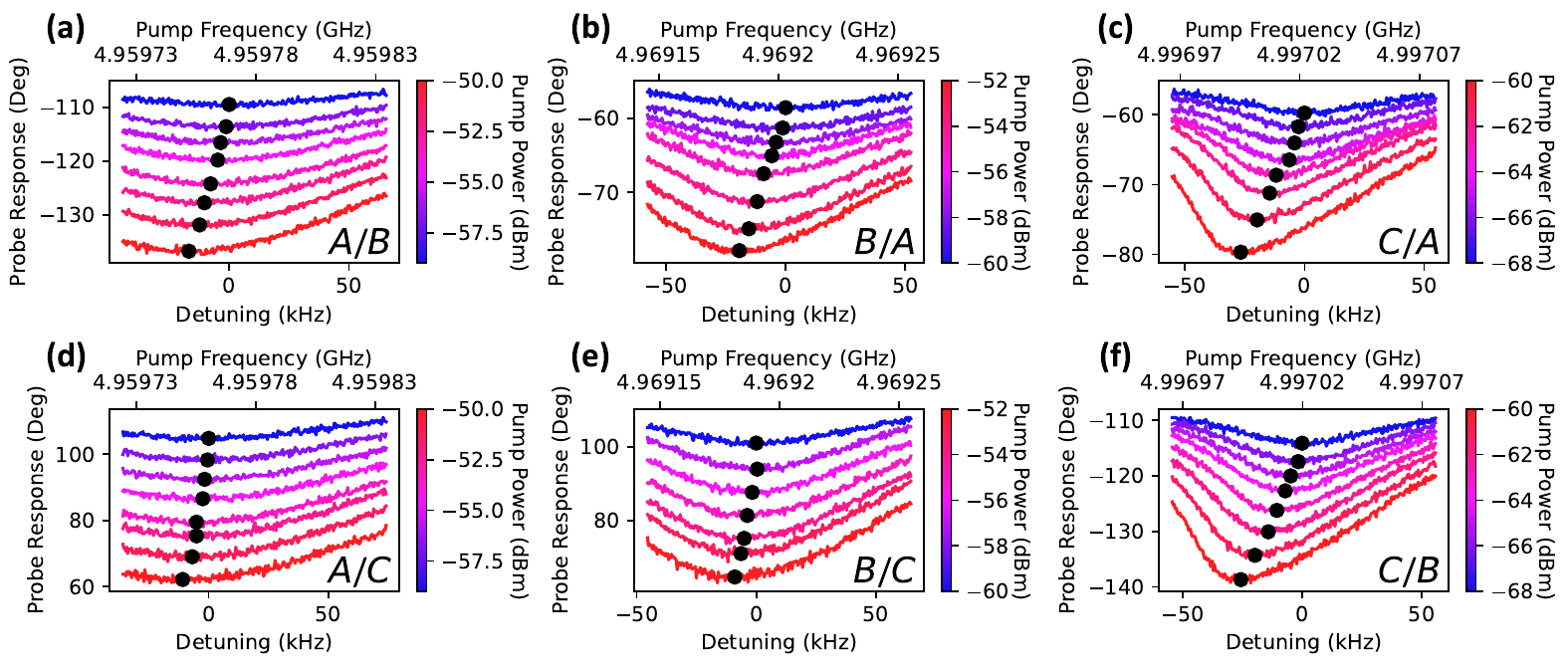}
	% \vspace{-0.6cm}
    \vspace{-0.2cm}
    \caption{\label{fig:raw_kerr_supp} 
    \textit{Self-Kerr and cross-Kerr shifts} (a)-(f) Two-tone spectroscopy of for all possible drive and monitor mode pairs. The combination shown in each subpanel is indicated by the label in the lower right: drive mode (A,B,C)/monitor mode (A,B,C). 
    %The measurement is repeated for eight different pump powers to observe the self-Kerr and cross-Kerr shifts per unit power. 
    Sample data for eight different drive powers are shown. Line color indicates the applied drive power.
    The Kerr-shifted frequencies of the drive and monitor modes are found by fitting the response function of a Kerr-shifted oscillator and identifying the minimum of the fitted function, which is marked with a black dot. The pump frequency where the minimum occurs ($x$-coordinate of the black dot) determines the self-Kerr-shifted frequency of the drive mode and the magnitude of the probe response ($y$-coordinate of the black dot) determines the cross-Kerr-shifted frequency of the monitor mode. 
    %The cross-Kerr shift on the monitor mode is determined by comparing the peak probe signal response to the measured frequency response of the mode. 
    The frequencies of the three drive/monitor modes, and their locations within the bands of the lattice are shown in Fig.~\ref{fig:mode_spectroscopy}(b).} 
  
\end{figure*}

By finding the pump frequency which produces the largest phase response for a given power, the effective drive mode frequency $\omega_D'$ can be identified.
To accurately identify this pump frequency and consequently $\omega_D'$, the measured phase shift is converted to the frequency of the monitor mode using the known phase response function identified in the initial transmission measurements described above.
% The method through which the frequency of a sensor mode can be tracked in two-tone spectroscopy is described in full in Sec.~\ref{app:TwoTone}.
The expected response function of a Kerr oscillator, given by Eq.~(\ref{eq:semiclassical_Kerr}), is then fitted to the transformed data and the absolute minimum of the function is identified.
The pump frequency where this minimum occurs returns $\omega_D'$, and the value of this minimum can be used to extract $\omega_M'$.
See Section~\ref{app:TwoTone} for more detail on the method we use to track the frequencies of sensor and target features in two-tone spectroscopy.
The two-tone spectroscopy configuration for carrying out the Kerr measurement is illustrated in Fig.~\ref{fig:two_tone}(d) and sample data for the measurement are shown in Fig.~\ref{fig:raw_kerr_supp}.

% Goals of section: 

% 1. Re-describe measurement protocol using pump/probe terminology from two-tone spectroscopy. 

% 3. Use extra space to emphasize that the technique doesn't require mode to be visible in transmission (with slight modification). That's why we're presenting this method instead of looking at self-Kerr shifts in transmission (or dispersive shift directly?).  x 

% 4. Describe data fit in more detail than main text

\begin{figure*}[t]
\centering
    \includegraphics[width=1\textwidth]{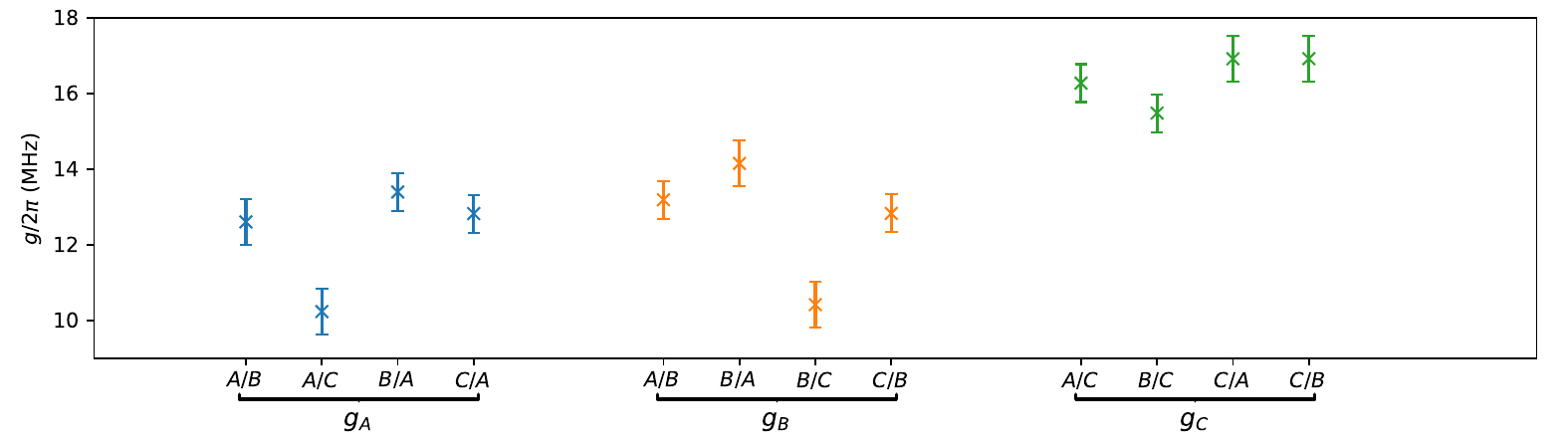}
	% \vspace{-0.6cm}
    \vspace{-0.2cm}
    \caption{\label{fig:pair_comparison} 
    \textit{Predicted qubit-mode couplings using different mode pairs.} Comparison of qubit-mode couplings returned by our measurement protocol for all six possible drive/monitor combinations of lattice modes $A$, $B$, and $C$. A summary of these extracted values is also shown in Table~\ref{app:gtable}.} 
  
\end{figure*}

\section{Mode Pair Comparisons}\label{app:3Modes}

In this section, we demonstrate the protocol on three lattice modes using all six possible drive/monitor mode pairings to verify that the predicted qubit-mode coupling is independent of the mode pair used for the measurement.
We label the modes selected for this comparison as $A$, $B$, and $C$, with frequencies $\omega_A/2\pi = 4.960$~GHz, $\omega_B/2\pi=4.969$~GHz, and $\omega_C/2\pi=4.997$~GHz,  respectively, and the transmon qubit is tuned to the frequency $\omega_q/2\pi = 4.593$~GHz.
The selected modes are indicated within the spectrum of the device displayed in Fig.~\ref{fig:mode_spectroscopy}(a).
Taking every combination, this process returns four separate coupling predictions for each individual mode: two using measurements of the self-Kerr shift, when it acts as the drive mode, and two using measurements of the cross-Kerr shift, when it acts as the monitor mode.
The predicted qubit-mode couplings returned by the protocol for these drive/monitor pairings are presented in Table~\ref{app:gtable} and shown graphically in Fig.~\ref{fig:pair_comparison}.
Error bars for the couplings are determined from the repeatability measurement described in Section~\ref{app:stability}.
Agreement between these predictions further validates the output of our protocol.

For each mode pair, an AC-Stark-shift measurement and Kerr-shift measurement are carried out to measure the frequency shifts per unit power (see Section~\ref{app:Protocol_Details}).
Two-tone spectroscopy data of the AC-Stark shift using all three possible choices of drive mode are presented in Fig.~\ref{fig:stark_full}.
Though AC-Stark data was collected for each possible pair combination to calculate the qubit-mode coupling, only three of the sets are displayed as the choice of monitor mode has no effect on this measurement.
The full set of Kerr-shift two-tone spectroscopy measurements are shown in Fig.~\ref{fig:raw_kerr_supp}.

\begin{table*}
\caption{\label{app:gtable}%
\textit{Predicted qubit-mode couplings using different mode pairs.} Comparison of the qubit-mode couplings measured through the protocol using all possible drive/monitor pairs of modes $A$, $B$, and $C$. A graphical depiction of these values, including systematic errors bars is shown in Fig.~\ref{fig:pair_comparison}.}
\def\arraystretch{1.3}%
% \begin{tabular}{>{\centering\arraybackslash}p{0.19\textwidth-2\tabcolsep}|>{\centering\arraybackslash}p{0.269\textwidth-2\tabcolsep}|>{\centering\arraybackslash}p{0.269\textwidth-2\tabcolsep}|>{\centering\arraybackslash}p{0.269\textwidth-2\tabcolsep}}
% \begin{tabular}{ p{0.19in} | p{0.269in} | p{0.269in}| p{0.269in}}
% \begin{tabular}{ |p{3cm}||p{3cm}|p{3cm}|p{3cm}|  }
\begin{tabular}{|c|c|c|c|}
\hline
Drive/Monitor Pair & $g_A/2\pi$~(MHz) & $g_B/2 \pi$~(MHz) & $g_C/2\pi$~(MHz)\\
\hline
$A$/$B$ & 12.6 & 13.2 & $\times$ \\
\hline
$B$/$A$ & 13.4 & 14.2 & $\times$\\
\hline
$A$/$C$ & 10.2 & $\times$ & 16.3\\
\hline
$C$/$A$ & 12.8 & $\times$ & 16.9\\
\hline
$B$/$C$ & $\times$ & 10.4 & 15.5\\
\hline
$C$/$B$ & $\times$ & 12.8 & 16.9\\
\hline
\end{tabular}
\end{table*}

%>{\centering\arraybackslash}p{0.1\textwidth-2\tabcolsep}|
% \begin{table*}
% \caption{\label{app:paramtable}%
% \textit{Device Design Parameters.} Comparison of numerically simulated device parameters and measured values for the inter-CPW hopping, transmons, and qubit-resonator coupling.  }
% %\begin{ruledtabular}
% \def\arraystretch{1.2}%
% \setlength{\tabcolsep}{4pt}
% \begin{tabular}{cccc}
% Resonator & $\vert t_{1} \vert/2\pi$ & $\vert t_{2} \vert/2\pi$\\
% \hline
% Finite Element & $47$ \mbox{MHz} & $94$ \mbox{MHz}\\
% DOS Comparison & $40$ MHz & $82$ \mbox{MHz} \\
% \hline
% Transmon & $E_C/ 2\pi$ & $E_{J,sum}/ 2\pi$\\
% \hline
% Finite Element & $165$ \mbox{MHz} & - \\
% \sixteen ~(Measured) & $125$ \mbox{MHz} & $100$ \mbox{GHz} \\
% \five ~(Measured) & $113$ \mbox{MHz} & $104$ \mbox{GHz} \\
% \eight ~(Measured) & $122$ \mbox{MHz} & $103$ \mbox{GHz} \\
% \hline
% Qubit-Resonator Coupling & $g_1/ 2\pi$ & $g_2/ 2\pi$ \\
% \hline
% Finite Element & $76$ \mbox{MHz} & $152$ \mbox{MHz} \\
% Measurement & $82.5$ \mbox{MHz} & $165$ \mbox{MHz} \\
% \end{tabular}
% \end{ruledtabular}
% \end{table*}

% Goals of section:

% 1. Explain that we want to verify that qubit-mode couplings are independent of the mode pair used

% 2. Describe modes used and all potential combos

% 3. Insert table and fig

% 4. Point to agreement in predictions using different pairs.

% In theory, should be able to fit whole 2D thing. Verification that with the right model (including phase drift) it fits to the whole thing. Good agreement between Kerr calculation and data set with that assumption

\section{Measurement Reproducibility}\label{app:stability}

\begin{figure*}[t]
\centering
    \includegraphics[width=1\textwidth]{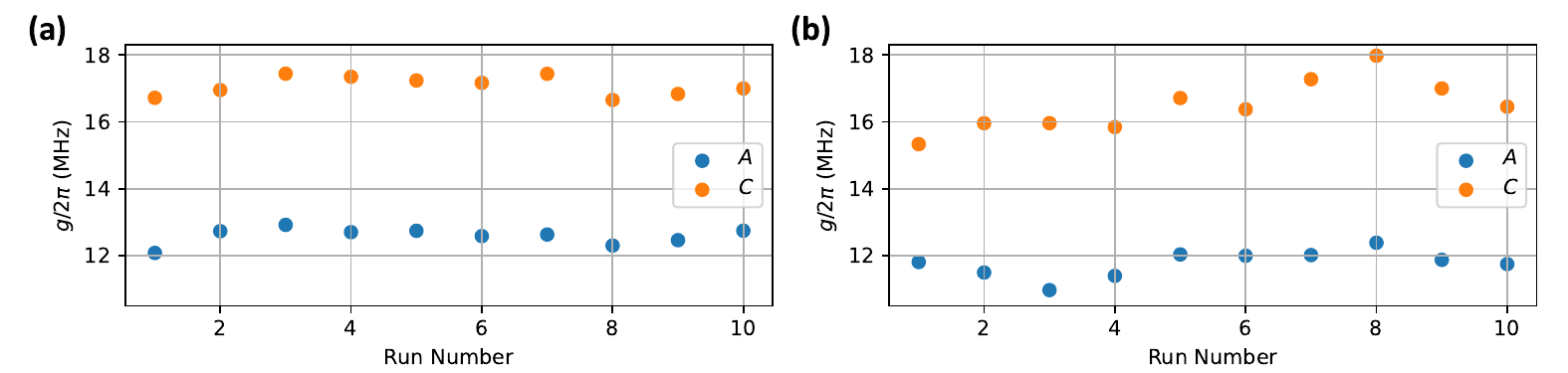}
	% \vspace{-0.6cm}
    \vspace{-0.2cm}
    \caption{\label{fig:stability} 
    \textit{Repeatability measurements.} Predicted qubit-mode couplings from ten repeated applications of the protocol over 27 hours using mode $C$ as the drive mode and mode $A$ as the monitor at two fixed qubit detunings: (a) with the qubit tuned to $\omega_q/2 \pi = 4.60$~GHz and (b) with the qubit tuned to $\omega_q/2 \pi = 4.18$~GHz.} 
  
\end{figure*}

To determine the reproducibility of the qubit-mode couplings returned by the measurement protocol, we apply the protocol ten times over 27 hours and characterize the spread of the predicted couplings.
For this process, we use mode $C$ as the drive mode and mode $A$ as the monitor. 
Since the magnitude of the AC-Stark and Kerr shifts per unit power decrease as the detuning increases, the spread of predicted couplings is expected to be dependent on the qubit frequency.
To account for this detuning dependence, we  carry out repeated measurements at two different qubit detunings, one with the qubit closer to the modes at $\omega_q/2\pi = 4.60$~GHz and one with the qubit further from the modes at $\omega_q/2\pi = 4.18$~GHz.
The couplings predicted by these measurements, using mode 
C as the drive mode and mode A as the monitor mode, are shown in Fig.~\ref{fig:stability}.

In these measurements, systematic variation in the predicted couplings was found to dominate over statistical error.
The values of $g_C$ and $g_A$ returned by the protocol are highly correlated, most likely due to qubit drift that occurs over the course of the measurement.
The magnitude of the drift observed in these measurements is used to determine error bars for the qubit-mode couplings returned by the protocol.
The standard deviations for the sets of couplings are found to be $600$~kHz for $g_C$, which is taken to be the error for predictions of drive-mode coupling, and $500$~kHz for $g_A$, which is taken to be the error for predictions of monitor-mode coupling.

\section{Semiclassical Solution to a Driven-Damped Kerr Resonator}
\label{app:semiclassic}

% \ak{Yuxin, can you add a topic sentence?}
In this section, we summarize the semiclassical solution for a driven dissipative Kerr resonator. The spectroscopy measurements described in the main text are fit to the theoretical predictions in this section to extract the Kerr coefficients.

We consider a driven damped Kerr oscillator, whose dynamics can be described by a Lindbladian master equation as 
\begin{align}
\label{neq:meq.lindblad.gen}
% \partial _{t} \hat \rho = - i [\hat H _{\text{tot}} , \hat \rho ] + \kappa  \mathcal{D} [ {\hat a} ] \hat \rho 
\partial _{t}  \rho = - i [ H _{\text{tot}} , \rho ] + \kappa  \mathcal{D} [ {  a} ]  \rho 
. 
\end{align}
Here we assume a rotating frame defined with respect to the external drive frequency $\omega _{\text{dr}}$, so that the Hamiltonian can be written as 
\begin{align}
% \hat H _{\text{tot}} = - \Delta _{\mathrm{dr}} {\hat a} ^\dag {\hat a} -\frac{\eta _{a}}{2} {\hat a} ^\dag {\hat a} ^\dag {\hat a}  {\hat a} + \left ( \mathcal{E} _{\mathrm{dr} }  {\hat a} ^\dag + \text{H.c.}
% \right ) 
H _{\text{tot}} = - \Delta _{\mathrm{dr}} { a} ^\dag { a} -\frac{\eta _{a}}{2} {  a} ^\dag { a} ^\dag {  a}  { a} + \left ( \mathcal{E} _{\mathrm{dr} }  {  a} ^\dag + \text{H.c.}
\right ) 
, 
\end{align}
with the detuning $\Delta _{\mathrm{dr}} = \omega _{\text{dr}}- \omega _{q}$. 
We can write the quantum Langevin equation for the mode operator as
\begin{align}
\partial _{t} {  a} = \left [- i  (\Delta _{\mathrm{dr}} + \eta _{a} {  a} ^\dag {  a}  ) - \frac{\kappa }{2} \right ]{  a} + i \mathcal{E} _{\mathrm{dr} }  + \sqrt{\kappa}  {  a} _{\mathrm{in}}
.
\end{align}
For a typical experiment, we have $\kappa/2 \pi \sim 100 ~$kHz and $\eta _{a}/2 \pi \sim 10~$Hz.

The semiclassical limit of Eq.~\eqref{neq:meq.lindblad.gen} has been extensively studied in the context of optical bistability (see references in the Introduction of Ref.~\cite{Walls1980}). 
Note also that the steady state of a driven-dissipative Kerr oscillator in Eq.~\eqref{neq:meq.lindblad.gen} can be solved analytically~\cite{Walls1980,Ciuti2016,Clerk2020}; however, such an exact solution involves special functions and is typically nontrivial to evaluate numerically.

As a first step, we can treat the system dynamics with a semiclassical approximation, so that the system state is fully characterized by a classical amplitude $\alpha$. More specifically, the intracavity amplitude is related to the drive amplitude via 
\begin{align}
\label{neq:eom.ss.alpha}
\left [ i  (\Delta _{\mathrm{dr}} + \eta _{a} |\alpha| ^{2} ) + \frac{\kappa }{2} \right ] \alpha  = i \mathcal{E} _{\mathrm{dr} } 
, 
\end{align}
and we can solve for the average photon number $\bar{n} _{a} = \langle {\hat a} ^\dag {\hat a} \rangle = |\alpha| ^{2} $
% $n _{\mathrm{qb}} = |\alpha| ^{2}$ 
via the following equation (cf., Eq.~(2.10) of~\cite{Walls1980}):
\begin{align}
\label{neq:nph.avg.cl}
\left [ (\Delta _{\mathrm{dr}} + \eta _{a} |\alpha| ^{2} ) ^{2} + \left ( \frac{\kappa }{2} \right ) ^{2} \right ] |\alpha| ^{2}  = |\mathcal{E} _{\mathrm{dr} } | ^{2} 
.
\end{align}
% Thus, the average photon number of the linearized dynamics can be computed as 
% \begin{align}
% \langle {\hat a} ^\dag {\hat a} \rangle = |\alpha| ^{2} 
% \end{align}
For a given drive amplitude $\mathcal{E} _{\mathrm{dr} }$ and detuning $\Delta _{\mathrm{dr}} $, Eq.~\eqref{neq:nph.avg.cl} defines an implicit function for the average photon number $\bar{n} _{a} = |\alpha| ^{2} $ under the semiclassical approximation:
\begin{align}
\label{neq:nph.avg.cl.na}
\left [ \left ( \bar{n} _{a} + \frac{\Delta _{\mathrm{dr}}}{\eta _{a} }\right) ^{2} + \left ( \frac{\kappa }{2 \eta _{a} } \right ) ^{2} \right ] \bar{n} _{a} = \frac{|\mathcal{E} _{\mathrm{dr} } | ^{2} }{\eta _{a} ^{2}}
. 
\end{align}
Equivalently, we can explicitly compute the detuning required to obtain steady state photon number $\bar{n} _{a} = |\alpha| ^{2} $ in the semiclassical limit as 
\begin{align}
\label{neq:nph.det.cl}
\Delta _{\mathrm{dr}} = - \eta _{a} \bar{n} _{a}  \pm \sqrt{\frac{|\mathcal{E} _{\mathrm{dr} } | ^{2} }{\bar{n} _{a}  } - \left ( \frac{\kappa }{2} \right ) ^{2} }
. 
\end{align}
One can also show that for the semiclassical equation to have multiple real steady-state solutions, the minimal drive amplitude is given by
\begin{align}
% \bar{a} _{\mathrm{crit}} = 
|\mathcal{E} _{\mathrm{dr,crit} }| = \sqrt{\frac{\kappa ^{3}}{3 ^{\frac{3}{2}}\eta _{a} }}
. 
\end{align}
Moreover, the maximal intracavity photon number is realized when the radical in Eq.~\eqref{neq:nph.det.cl} vanishes, which corresponds to the set of parameters
\begin{align}
\bar{n} _{a} =
|\alpha| ^{2} = \frac{|\mathcal{E} _{\mathrm{dr} } | ^{2} }{\left ( \frac{\kappa }{2} \right ) ^{2} }, \quad 
\Delta _{\mathrm{dr}} = - \frac{\eta _{a} }{\left ( \frac{\kappa }{2} \right ) ^{2} } |\mathcal{E} _{\mathrm{dr} } | ^{2} 
. 
\end{align}
The peak value intracavity photon number with the critical drive power is thus given by
\begin{align}
\bar{n} _{a,\mathrm{crit}} =
\frac{|\mathcal{E} _{\mathrm{dr,crit} } | ^{2} }{\left ( \frac{\kappa }{2} \right ) ^{2} }
= \frac{4 \kappa  }{27 \eta _{a}}
. 
\end{align}
When the intracavity photon number is much greater than this critical value, the semiclassical approximation breaks down, in which case we should solve the full quantum master equation Eq.~\eqref{neq:meq.lindblad.gen}.

The experiment measures the intensity and the phase of the transmitted signal. We can use the input-output relation to compute the transmitted field as
\begin{align}
\alpha _{\text{out}} 
= \kappa _{\mathrm{out} } \alpha = \frac{i \sqrt{\kappa _{\mathrm{out} }} \mathcal{E} _{\mathrm{dr} } }{\frac{\kappa }{2} + i  (\Delta _{\mathrm{dr}} + \eta _{a} |\alpha| ^{2} ) }
, 
\end{align}
where the intracavity photon number is given by the real-valued solution to Eq.~\eqref{neq:nph.avg.cl.na}, which is a cubic equation in $\bar{n} _{a} $:
\begin{align}
\left [ \left ( \bar{n} _{a} + \frac{\Delta _{\mathrm{dr}}}{\eta _{a} }\right) ^{2} + \left ( \frac{\kappa }{2 \eta _{a} } \right ) ^{2} \right ] \bar{n} _{a} = \frac{|\mathcal{E} _{\mathrm{dr} } | ^{2} }{\eta _{a} ^{2}}
. 
\end{align}
When the photon number is sufficiently small such that the following condition holds
\begin{align}
\eta _{a} \bar{n} _{a} \ll |\Delta _{\mathrm{dr}} |, \frac{\kappa }{2}
,
\end{align}
the cavity response can be further approximated by linear response 
\begin{align}
\bar{n} _{a}  \simeq \frac{|\mathcal{E} _{\mathrm{dr} } | ^{2} }{\Delta _{\mathrm{dr}}  ^{2}+ \left ( \frac{\kappa }{2} \right ) ^{2}}
, 
\end{align}
which in turn corresponds to the following constraint on the drive power:
\begin{align}
\eta _{a} \frac{|\mathcal{E} _{\mathrm{dr} } | ^{2} }{\Delta _{\mathrm{dr}}  ^{2}+ \left ( \frac{\kappa }{2} \right ) ^{2}}  \ll \left ( \frac{\kappa }{2} \right ) ^{2}
% \sqrt{\Delta _{\mathrm{dr}}  ^{2}+ \left ( \frac{\kappa }{2} \right ) ^{2}}
\Rightarrow |\mathcal{E} _{\mathrm{dr} } | ^{2} \ll \frac{\frac{\kappa }{2}  \left [ \Delta _{\mathrm{dr}}  ^{2}+ \left ( \frac{\kappa }{2} \right ) ^{2}\right ]  }{\eta _{a} }
. 
\end{align}
In experimental measurements, we are typically below the critical driving power (i.e., no bistability), in which case we can compute the intracavity photon number perturbatively:
\begin{align}\label{eq:semiclassical_Kerr}
& \bar{n} _{a} = \frac{|\mathcal{E} _{\mathrm{dr} } | ^{2} }{\eta _{a} ^{2} \left [ \left ( \bar{n} _{a} + \frac{\Delta _{\mathrm{dr}}}{\eta _{a} }\right) ^{2} + \left ( \frac{\kappa }{2 \eta _{a} } \right ) ^{2} \right ] } 
% \\\simeq & 
% \frac{|\mathcal{E} _{\mathrm{dr} } | ^{2} }{ \Delta _{\mathrm{dr}} ^{2} + \frac{\kappa ^{2} }{4  }} \left [ 1 - \frac{2 \eta _{a} \Delta _{\mathrm{dr}} 
%  |\mathcal{E} _{\mathrm{dr} } | ^{2} }{\left (  \Delta _{\mathrm{dr}} ^{2} + \frac{\kappa ^{2} }{4  } \right ) ^{2}} - \frac{\eta _{a} ^{2} |\mathcal{E} _{\mathrm{dr} } | ^{4} }{\left (  \Delta _{\mathrm{dr}} ^{2} + \frac{\kappa ^{2} }{4  } \right ) ^{3}}   \right ]
. 
\end{align}
The cross-Kerr shifts measured in the experiments are proportional to the intracavity photon number of the driven mode. Fitting the cross-Kerr shifts to Eq.~\eqref{eq:semiclassical_Kerr} (assuming an unknown conversion coefficients), we can extract the self-Kerr coefficient of the driven mode. The details of this fit are described in the next section.

\begin{figure*}[t]
\centering
    \includegraphics[width=1\textwidth]{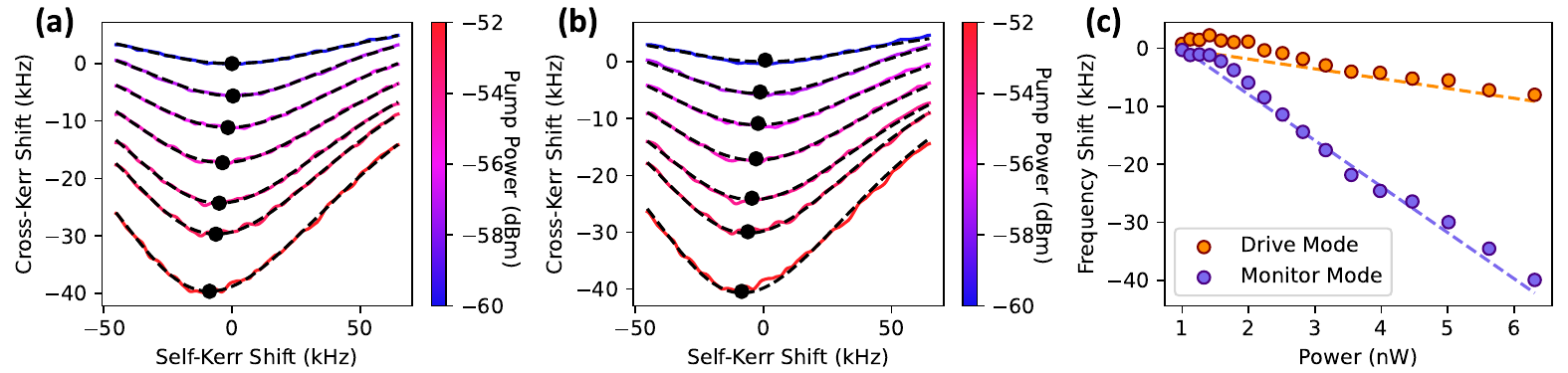}
	% \vspace{-0.6cm}
    \vspace{-0.2cm}
    \caption{\label{fig:fit_supp} 
    \textit{Comparison of fit methods.} Applications of the 1D and 2D fit methods to a Kerr measurement using mode $C$ as the drive mode and mode $B$ as the monitor mode. A Gaussian filter is applied to the data prior fitting to reduce the impact of noise of the more delicate fits. (a) Filtered data for seven applied drive powers along with the best-fit curves (black dashed lines) to Eq.~(\ref{eq:semiclassical_Kerr}) at each individual drive power, showing the quality of fit for the 1D fit method. Fitted minima indicated by black dots.
    (b) Filtered data for seven applied drive powers along with the best fit from the 2D fit (black dashed lines) to all drive powers simultaneously and including slow phase drift. Fitted minima indicated by black dots.
    (c) Comparison of extracted Kerr shifts versus drive power from the two fit methods. Colored circles indicate the extracted frequency shifts of the drive and monitor modes, as determined from the 1D fit method. Dashed lines indicate the extracted frequency shifts from the 2D fit, which are required to be linear by construction. 
    %
    % The response function of a driven Kerr oscillator [Eq.~(\ref{eq:semiclassical_Kerr})] is fit to the measured frequency response at each drive power separately using the 1D fit method. The fitted functions are shown as dashed black lines with minima indicated by a black dot. 
    % (b) The response function of a driven Kerr oscillator is fit to the measured frequency response for all drive powers simultaneously using the 2D fit method. The fitted function is represented by a dashed black line for each drive power with the minimum indicated by a black dot. (c) Colored circles indicate the extracted frequency shifts of the drive and monitor modes, as determined from the 1D fit method, versus the drive power. The dashed lines indicate the extracted frequency shifts from the 2D fit method versus the drive power. The simultaneous fit constrains the fitted self-Kerr and cross-Kerr shifts to be linear with respect to power.
    } 
  
\end{figure*}

\section{Kerr Measurement Fits}\label{app:2Dfit}

In the Kerr-shift portion of our measurement protocol, we fit the expected response of a Kerr oscillator [Eq.~(\ref{eq:semiclassical_Kerr})] derived in Section~\ref{app:semiclassic} to the measured response of the drive mode in order to track the self-Kerr and cross-Kerr shifts on the frequencies of \emph{both} the drive and monitor modes.
In this section, we discuss two procedures for carrying out this fit that serve different purposes: a ``1D'' fit, where the model, derived in Section~\ref{app:semiclassic}, is separately fit to the measured response versus drive frequency for each individual drive power, and a ``2D'' fit, where the model is fit to the full set of measured responses for all drive powers and drive frequencies at once.
Empirically, we find that the 1D fit is robust to experimental imperfections but does not yield a single set of fit parameters that can describe the data at all drive powers.
Conversely, the 2D fit, by construction, finds a single set of mode parameters that describes the dataset as a whole, but requires incorporation of additional empirical fit parameters in order to converge. 
%Conversely, the 2D fit finds a set of parameters that describes the whole system but is more sensitive.

In the 1D fit, Eq.~(\ref{eq:semiclassical_Kerr}) is fit to the measurements at each drive power individually. The effective drive power in the mode $\vert\mathcal{E} _{\mathrm{dr}}\vert^2$ is determined independently for each power, along with the frequency, linewidth, and self-Kerr coefficient of the drive mode.
% the drive power $\vert\mathcal{E} _{\mathrm{dr}}\vert^2$ is fit separately for each trace

The 2D fit incorporates the fact that these parameters are not independent and the fact that all of the data sets originate from a single value of the drive mode frequency, linewidth, and self-Kerr coefficient, along with a single conversion factor 
$\vert\mathcal{E} _{\mathrm{dr}}\vert^2 = \zeta P$ between the effective intracavity drive power and the applied laboratory power.
We find that the unified 2D fit can capture the Kerr data sets as a whole as long as one additional effect is introduced to the model: slow drift in the phase of the probe tone.
This drift is empirically observed in the system and introduces systematic error to the measured cross-Kerr shift on the frequency of the monitor.
To account for slow phase drift in the fit, we incorporate an independent phase offset \emph{for each} drive power, and the fit determines these in addition to the fundamental cavity parameters above.
A comparison between the 1D and 2D fits for a Kerr measurement using mode $C$ as the drive and mode $B$ as the monitor is shown in Fig.~\ref{fig:fit_supp}.

%To account for phase drift in the model, a phase drift parameter is introduced for each value of the drive power.
%This parameter is associated with a constant offset in the measured response of the mode at that drive power.

After incorporating the effect of slow phase drift,  results of the 2D fit agree well with measured data, which supports the validity of the simple single-mode Kerr oscillator model for the response of individual modes in our multimode system. 
However, as the 1D method reliably fits the minima of the measured mode response without requiring additional free parameters, the data in the main text and other sections of the Supplemental Material are solely fit through the 1D method.

\section{\(O_{\aLabel\aLabel}\) and \(O_{\aLabel\bLabel}\) for the Transmon}\label{app:transmon_operators}

In the end matter, we computed the fourth-order correction \eq{transmonFourthOrder} to the Hamiltonian for two modes coupled via a transmon qubit, and the self- and cross-Kerr shifts \(\chi_{\aLabel\aLabel}\) and \(\chi_{\aLabel\bLabel}\) for a transmon in the ground state. Here we show the full shift operators \(O_{\aLabel\aLabel}\) and \(O_{\aLabel\bLabel}\) for arbitrary transmon excitation number, which act on the infinite-dimensional transmon Hilbert space. The shifts given in the end matter are \(\chi_{\aLabel\aLabel}=\bra 0 O_{\aLabel\aLabel} \ket 0\) and \(\chi_{\aLabel\bLabel} = \bra 0 O_{\aLabel\bLabel} \ket 0\).

The self-Kerr operator on the entire transmon subspace is
\begin{align}
    O_{\aLabel\aLabel} = g_\aLabel^4 (c^\dagger c^\dagger \xi_{\aLabel\aLabel,2} c c+c^\dagger\xi_{\aLabel\aLabel,1} c+\xi_{\aLabel\aLabel,0}),
\end{align}
where [the operators \(f_{i,\mu}\) and \(h_{ij,\mu}\) are defined in Eqs.~(\ref{eq:f_def},\,\ref{eq:h_def}) of the main text]
\begin{align}
    \xi_{\aLabel\aLabel,0}
        &= 2f_{\aLabel,0}^3 - 3f_{\aLabel,0}^2f_{\aLabel,1} + f_{\aLabel,0}f_{\aLabel,1}^2 + (-f_{\aLabel,0}^2 + 2f_{\aLabel,0}f_{\aLabel,1} - f_{\aLabel,1}^2)h_{\aLabel\aLabel,0},
    \nonumber\\
    \xi_{\aLabel\aLabel,1}
        &= -2f_{\aLabel,0}^3 + 4f_{\aLabel,0}^2f_{\aLabel,1} - 4f_{\aLabel,0}f_{\aLabel,1}^2 + 6f_{\aLabel,1}^3 - 6f_{\aLabel,1}^2f_{\aLabel,2} + 2f_{\aLabel,1}f_{\aLabel,2}^2 + (-2f_{\aLabel,1}^2 + 4f_{\aLabel,1}f_{\aLabel,2} - 2f_{\aLabel,2}^2)h_{\aLabel\aLabel,1},
    \nonumber\\
    \xi_{\aLabel\aLabel,2}
        &=
        \frac 1 2 [
            -f_{\aLabel,0}^2f_{\aLabel,1} + 3f_{\aLabel,0}f_{\aLabel,1}^2 - 4f_{\aLabel,1}^3 + 4f_{\aLabel,1}^2f_{\aLabel,2} - 4f_{\aLabel,1}f_{\aLabel,2}^2 + 4f_{\aLabel,2}^3 - 3f_{\aLabel,2}^2f_{\aLabel,3} + f_{\aLabel,2}f_{\aLabel,3}^2 
            \nonumber\\
            &\qquad+ (f_{\aLabel,0}^2 - 2f_{\aLabel,0}f_{\aLabel,1} + f_{\aLabel,1}^2)h_{\aLabel\aLabel,0} 
            + (-f_{\aLabel,2}^2 + 2f_{\aLabel,2}f_{\aLabel,3} - f_{\aLabel,3}^2)h_{\aLabel\aLabel,2}
        ].
\end{align}
The matrix element for the first excited state is
\begin{align}
    \bra 1 O_{\aLabel\aLabel} \ket 1 = g_\aLabel^4(\xi_{\aLabel\aLabel,1}|_{c^\dagger c = 0}+\xi_{\aLabel\aLabel,0}|_{c^\dagger c = 1})
    = \frac{2 g_\aLabel^4 \alpha[\Delta_\aLabel^3+5(\alpha \Delta_\aLabel^2-\alpha^2 \Delta_\aLabel)+3\alpha^3]}
    {\Delta_\aLabel^3 (\Delta_\aLabel-\alpha)^3 (2\Delta_\aLabel-3\alpha)}.
    \label{eq:firstTransmonSK}
\end{align}
The cross-Kerr operator on the entire transmon subspace is
\begin{align}
    O_{\aLabel\bLabel} &= g_\aLabel^2 g_\bLabel^2(c^\dagger c^\dagger \xi_{2,\aLabel\bLabel} c c+c^\dagger \xi_{1,\aLabel\bLabel} c+\xi_{0,\aLabel\bLabel}),
\end{align}
where
\begin{align}
    \xi_{\aLabel\bLabel,0} &= \frac 1 2 [
        4f_{\aLabel,0}f_{\bLabel,0}^2 - 3f_{\aLabel,1}f_{\bLabel,0}^2 - 3f_{\aLabel,0}f_{\bLabel,0}f_{\bLabel,1} + f_{\aLabel,1}f_{\bLabel,0}f_{\bLabel,1} + f_{\aLabel,0}f_{\bLabel,1}^2 
        \nonumber\\
        &\qquad
        + (-f_{\aLabel,0}^2 + 2f_{\aLabel,0}f_{\aLabel,1} - f_{\aLabel,1}^2 - f_{\aLabel,0}f_{\bLabel,0} + f_{\aLabel,1}f_{\bLabel,0} + f_{\aLabel,0}f_{\bLabel,1} - f_{\aLabel,1}f_{\bLabel,1})h_{\aLabel\bLabel,0}
    ]+\{\aLabel\leftrightarrow \bLabel\},
    \nonumber\\
    \xi_{\aLabel\bLabel,1} &= 
    -2f_{\aLabel,0}f_{\bLabel,0}^2 + 2f_{\aLabel,1}f_{\bLabel,0}^2 + 2f_{\aLabel,0}f_{\bLabel,0}f_{\bLabel,1} - 2f_{\aLabel,1}f_{\bLabel,0}f_{\bLabel,1} - 2f_{\aLabel,0}f_{\bLabel,1}^2 + 6f_{\aLabel,1}f_{\bLabel,1}^2 - 3f_{\aLabel,2}f_{\bLabel,1}^2 
    \nonumber\\
    &\qquad
    - 3f_{\aLabel,1}f_{\bLabel,1}f_{\bLabel,2} + f_{\aLabel,2}f_{\bLabel,1}f_{\bLabel,2} + f_{\aLabel,1}f_{\bLabel,2}^2 + 
    \nonumber\\
    &\qquad
    (-f_{\aLabel,1}^2 + 2f_{\aLabel,1}f_{\aLabel,2} - f_{\aLabel,2}^2 - f_{\aLabel,1}f_{\bLabel,1} + f_{\aLabel,2}f_{\bLabel,1} + f_{\aLabel,1}f_{\bLabel,2} - f_{\aLabel,2}f_{\bLabel,2})h_{\aLabel\bLabel,1} + \{\aLabel\leftrightarrow \bLabel\},
    \nonumber\\
    \xi_{\aLabel\bLabel,2} &= \frac 1 4 [
        -f_{\aLabel,1}f_{\bLabel,0}^2 - f_{\aLabel,0}f_{\bLabel,0}f_{\bLabel,1} + 3f_{\aLabel,1}f_{\bLabel,0}f_{\bLabel,1} + 3f_{\aLabel,0}f_{\bLabel,1}^2 - 8f_{\aLabel,1}f_{\bLabel,1}^2 + 4f_{\aLabel,2}f_{\bLabel,1}^2 + 4f_{\aLabel,1}f_{\bLabel,1}f_{\bLabel,2} 
        \nonumber\\
        &\qquad
        - 4f_{\aLabel,2}f_{\bLabel,1}f_{\bLabel,2} - 4f_{\aLabel,1}f_{\bLabel,2}^2 + 8f_{\aLabel,2}f_{\bLabel,2}^2 - 3f_{\aLabel,3}f_{\bLabel,2}^2 - 3f_{\aLabel,2}f_{\bLabel,2}f_{\bLabel,3} + f_{\aLabel,3}f_{\bLabel,2}f_{\bLabel,3} + f_{\aLabel,2}f_{\bLabel,3}^2 
        \nonumber\\
        &\qquad
        + (f_{\aLabel,0}^2 - 2f_{\aLabel,0}f_{\aLabel,1} + f_{\aLabel,1}^2 + f_{\aLabel,0}f_{\bLabel,0} - f_{\aLabel,1}f_{\bLabel,0} - f_{\aLabel,0}f_{\bLabel,1} + f_{\aLabel,1}f_{\bLabel,1})h_{\aLabel\bLabel,0} 
        \nonumber\\
        &\qquad
        + (-f_{\aLabel,2}^2 + 2f_{\aLabel,2}f_{\aLabel,3} - f_{\aLabel,3}^2 - f_{\aLabel,2}f_{\bLabel,2} + f_{\aLabel,3}f_{\bLabel,2} + f_{\aLabel,2}f_{\bLabel,3} - f_{\aLabel,3}f_{\bLabel,3})h_{\aLabel\bLabel,2}
    ]+\{\aLabel\leftrightarrow \bLabel\}.
\end{align}
The matrix element for the first excited state is
\begin{align}
    \bra1 &O_{\aLabel\bLabel} \ket 1 =g_\aLabel^2 g_\bLabel^2(\xi_{\aLabel\bLabel,1}|_{c^\dagger c = 0}+\xi_{\aLabel\bLabel,0}|_{c^\dagger c = 1})
    \nonumber\\
    &= -\frac {2 g_\aLabel^2 g_\bLabel^2 \alpha
    [
        \Delta_\aLabel^2 \Delta_\bLabel^2 + 2 \alpha \Delta_\aLabel \Delta_\bLabel(\Delta_\aLabel + \Delta_\bLabel)
        -\alpha^2(8 \Delta_\aLabel \Delta_\bLabel + \Delta_\aLabel^2 + \Delta_\bLabel^2)
        +
        4 \alpha^3(\Delta_\aLabel + \Delta_\bLabel)
        -
        3 \alpha^4
    ]}
    {\Delta_\aLabel \Delta_\bLabel (\Delta_\aLabel-\alpha)^2 (\Delta_\bLabel-\alpha)^2 (\Delta_\aLabel + \Delta_\bLabel - 3\alpha)}
    \bigg( \frac 1 {\Delta_\aLabel} + \frac 1 {\Delta_\bLabel} \bigg).
    \label{eq:firstTransmonCK}
\end{align}
In the two-level system limit \(\alpha\to\infty\) the expressions in Eqs.~(\ref{eq:firstTransmonSK},\,\ref{eq:firstTransmonCK}) approach the \(\langle\sigma_z\rangle=1\)
case of Eqs.~(\ref{eq:TLSsk},\,\ref{eq:TLSck}). In the linear limit \(\alpha\to0\), \(f_{i,\mu}=1/\Delta_i\) and \(h_{ij,\mu}=1/(\Delta_i+\Delta_j)\), which gives \(O_{\aLabel\aLabel} = O_{\aLabel\bLabel} = 0\).

\end{document}